\documentclass{agujournal2019}
\usepackage{url} 
\usepackage{lineno}
\usepackage{multirow}
\usepackage[finalnew]{trackchanges} 
\usepackage{soul}
\usepackage{float}
\usepackage{natbib}
\usepackage{gensymb}
\draftfalse

\journalname{JGR: Space Physics}

\begin{document}

%
%


\title{Constraining Electron-Impact Ionization of O$_2$ Through UV Aurora Observations at Ganymede}
%
%




\authors{Stefan Duling\affil{1}, Joachim Saur\affil{1}, Darrell Strobel\affil{2}, Philippa Molyneux\affil{3}, Jamey R. Szalay\affil{4}, Thomas K. Greathouse\affil{3}}


\affiliation{1}{Institute of Geophysics and Meteorology, University of Cologne, Cologne, Germany}
\affiliation{2}{Department of Earth and Planetary Sciences, The Johns Hopkins University, Baltimore, MD, USA}
\affiliation{3}{Southwest Research Institute, San Antonio, TX, USA}
\affiliation{4}{Department of Astrophysical Sciences, Princeton University, Princeton, NJ, USA}







\begin{keypoints}
\item OI 1356 \r{A} emissions enable direct constraints on electron-impact ionization of O$_2$, with the ionization-to-excitation ratio confined to 10-60.
\item A map of ionization rates derived from 8-120 R emissions at Ganymede shows electron-impact is at least 10 times stronger than photoionization.
\item We find a global O$_2$ ionization rate of $1.3-7.6\times10^{26}$ s$^{-1}$, indicating 0.5-11 kg s$^{-1}$ of ionospheric O$_2^+$ outflow and 0.03-0.5 cm Myr$^{-1}$ of surface erosion.
\end{keypoints}

%
%

%
%


\begin{abstract}
While photoionization rates of Ganymede's O$_2$ dominated atmosphere are well constrained, the contribution of electron-impact ionization is rather uncertain.
Previous quantitative estimates have relied on assumptions about densities and energy distributions of precipitating electrons, or on rare spacecraft measurements that cannot be unambiguously mapped to the regions of ionization.
In this study, we present a novel approach to quantify electron-impact ionization rates directly through OI 1356 \r{A} emission brightness observations.
The analysis of measured cross sections reveals that the ionization-to-excitation ratio is limited to 10-60 over all electron energies, reducing the uncertainty of estimating ionization rates to a factor less than 6.
We apply this method to Juno UVS observations of Ganymede's aurora.
We find that the OI 1356 \r{A} brightness of the auroral ovals is well described by 3-5° latitude wide Gaussian distributions centered on the open-closed field line boundary, with an average peak of 120 R.
The average brightness outside the ovals in the polar and equatorial background regions is $\sim$8 R.
From these observations, we derive a global map of electron-impact ionization rates, which are at least an order of magnitude higher than photoionization rates.
The estimated total global ionization rate is $1.3-7.6\times10^{26}$ s$^{-1}$, with average column rates of $\sim5\times10^{9}$ cm$^{-2}$s$^{-1}$ in the ovals and $\sim3\times10^{8}$ cm$^{-2}$s$^{-1}$ in the background regions.
Comparison of radio occultation measurements with predicted electron densities indicates that transport processes are the dominant loss mechanism in Ganymede's ionosphere.
The rate of ionospheric outflow of O$_2^+$ is $0.1-2\times10^{26}\;\mathrm{s}^{-1}$ or $0.5-11\;\mathrm{kg}\;\mathrm{s}^{-1}$, indicating $0.03-0.5$ cm Myr$^{-1}$ erosion of Ganymede's surface ice.
\end{abstract}

\section*{Plain Language Summary}
Ganymede, Jupiter's largest moon, has a thin atmosphere made mostly of oxygen.
When energetic particles from space collide with these oxygen molecules, they can create charged ions.
It is difficult to estimate how often these ionization interactions happen because oxygen density and, especially, precipitating particle fluxes have hardly been measured yet.
In this study, we introduce a new way to estimate ionization rates directly from brightness measurements of ultraviolet auroral glow of an oxygen atmosphere.
This method takes advantage of the well-known ratio of auroral brightness to ionization rates, which is relatively constant (10-60) and almost independent of the precipitating particle flux.
By analyzing data from NASA's Juno spacecraft, we create a simple global map for the auroral glow and associated ionization rates.
We compare predicted oxygen ion densities with measurements and conclude that most ions are either lost at Ganymede's surface or to space, creating a steady flow of material from Ganymede into its environment.

%
%

%


%
%
%
%

\section{Introduction}
\label{sec_introduction}

The space environment of Ganymede, Jupiter's largest satellite, is one of the most complex in the solar system.
Ganymede possesses a dynamo-generated magnetic field embedded within Jupiter's gigantic, plasma-filled magnetosphere.
Various processes, such as sublimation, sputtering, and radiolysis, release particles from its predominantly water-ice surface, forming a tenuous atmosphere.
Electrons precipitating from Jupiter's and Ganymede's magnetospheres, guided by the intricate magnetic field topology, can interact with the atmospheric particles and trigger a range of physical processes.
Electron-impact ionization, alongside photoionization, produces charged particles that constitute an ionosphere.
Excitation interactions generate ultraviolet auroral emissions, which exhibit spatial and temporal variability.
In this study, we investigate electron-impact ionization rates and resulting properties of Ganymede's ionosphere by analyzing the brightness of its aurora.

As primary constituents of Ganymede's atmosphere, O$_2$ \citep{Hall1998} and H$_2$O \citep{Roth2021} molecules have been inferred from Hubble Space Telescope (HST) observations of auroral emissions from atomic oxygen.
Additionally, an extended but more dilute corona of atomic H was detected with the ultraviolet spectrometer (UVS) of the Galileo spacecraft \citep{Barth1997} and later confirmed with the Space Telescope Imaging Spectrograph (STIS) of HST \citep{Feldman2000,Alday2017}.
It is speculated that H originates from surface sputtering and dissociation of H$_2$O \citep{Barth1997} or from dissociation of atmospheric H$_2$ \citep{Alday2017,CarberryMogan2022}.
This would be consistent with models of Ganymede's atmosphere \citep{Marconi2007,Leblanc2017,Vorburger2024} that predict H$_2$O, O$_2$, H$_2$, H, O and OH molecules as products of sputtering, radiolysis and sublimation at the surface and reactions caused by interactions with solar photons and impacting electrons.
Of these molecules, O$_2$ and H$_2$ are the only species that do not stick to the surface upon re-impact due to their lower sublimation temperatures.
Thus, they have significant lifetimes of about 10 days, which is sufficient time to form an atmosphere \citep[e.g.][]{Vorburger2024}.
Measured surface temperatures ranging from 80 K on the night side to 150 K at the subsolar point \citep{Orton1996} suggest an O$_2$ scale height of approximately 20 km.
Models either reproduced this scale height \citep{Turc2014} or predicted larger numbers, up to 50 km \citep{Leblanc2017}.
Inferred O$_2$ column densities from HST observations of $10^{14}-10^{15} \;\mathrm{cm}^{-2}$ \citep{Hall1998} suggest a surface density of about $10^{8} \;\mathrm{cm}^{-3}$ \citep{Feldman2000}.
However, the lighter H$_2$ molecules reach much higher altitudes and can more readily escape.
They are modeled to be the dominant species at 100 km altitude and above ($10^{4}-10^{6}$ cm$^{-3}$) \citep{Turc2014}.
Although plasma bombardment and, consequently, sputtering are expected to be stronger at the polar caps, connected to the Jovian magnetosphere through open magnetic field lines, or near the boundary between open and closed field lines (OCFB), the abundances of O$_2$ and H$_2$ are modeled as more or less globally uniform due to their long lifetimes \citep[e.g.][]{Vorburger2024}.
In contrast, sublimation of H$_2$O on the day side leads to a locally enhanced, short-lived H$_2$O abundance, that can even dominate O$_2$ by one order of magnitude at the subsolar point \citep{Turc2014,Roth2021}.
However, additional observations of Ganymede's optical aurora detected only an upper limit of $3\times10^{13} \;\mathrm{cm}^{-2}$ for H$_2$O and are consistent with an O$_2$ dominated atmosphere \citep{Kleer2023,Milby2024}.
Similarly, recent near-infrared observations with the James Webb Space Telescope did also not detect an enhanced H$_2$O abundance \citep{BockeleeMorvan2024}.
Instead, these observations revealed the presence of CO$_2$ in Ganymede's atmosphere, primarily confined to the northern polar cap, with column densities up to $10^{14} \;\mathrm{cm}^{-2}$.

Ganymede's ionosphere is formed by ionization of the atmospheric neutrals.
Radio occultation observations are the only available direct measurements of the ionospheric plasma density up to now.
During Galileo's flybys eight radio occultation experiments were carried out.
While it is known that among these, there were one strong detection and two weak detections of an ionosphere \citep{McGrath2004}, only the two observations during the G8 flyby were published, measuring $2000 \pm 1500$ cm$^{-3}$ (ingress) and $5000 \pm 2000$ cm$^{-3}$ (egress) electrons in the lower ionosphere \citep{Kliore1998}.
In 2021, the radio occultation experiment during Juno's flyby provided values of $2000 \pm 500$ cm$^{-3}$ (ingress, detection) and $400\pm 500$ cm$^{-3}$ (egress, non-detection) for $\sim$10 km altitude \citep{Buccino2022}.
Additionally, \cite{Yasuda2024} used an occultation between the Jovian auroral radio sources and the Galileo spacecraft during the G1 flyby to constrain the maximum ionospheric density to 300 cm$^{-3}$ in the open field line region and 20 cm$^{-3}$ in the closed field line region.
The particle detection and plasma wave instruments onboard Galileo and Juno measured electron densities ranging from $5$ to $200 \;\mathrm{cm}^{-3}$ at the spacecrafts' locations \citep{Eviatar2001a,Frank1997,Gurnett1996,Williams1997a,Williams1997c,Allegrini2022a,Clark2022,Ebert2022a,Kurth2022a}.
However, none of the flyby trajectories crossed the altitude of a potential ionosphere directly.
The G2 flyby reached the lowest altitude of 263 km, where $200 \;\mathrm{cm}^{-3}$ were detected \citep{Eviatar2001a}.
Thus, extrapolations of in situ electron measurements to determine ionospheric densities \citep{Eviatar2001} are uncertain and not fully suited to constrain Ganymede's ionosphere in detail.

Efforts have been made to model Ganymede's ionosphere using simple chemical models \citep{Eviatar2001a,Cessateur2012,Waite2024} or 3D Monte Carlo approaches \citep[e.g.][]{Carnielli2019,Beth2025}.
The ionization rate is the most crucial parameter for populating these models with ions and electrons.
At the same time, ionization might constitute a significant loss for Ganymede's neutral atmosphere, according to model results \citep[e.g.][]{Marconi2007,Turc2014,Leblanc2017,Vorburger2024}.
While the photoionization rate at Ganymede is well constrained by knowledge of solar radiation, electron-impact ionization is less quantified but expected to contribute significantly as well \citep{CarberryMogan2023}.
The column electron-impact ionization rate of molecular oxygen is usually calculated by the product of an ionization frequency and a neutral column density, $\nu_\mathrm{ion}N_{\mathrm{O}_2}$.
Despite considerable modeling efforts, the abundance of oxygen remains uncertain by a factor of approximately 10.
More importantly, ionizing electron fluxes have rarely been measured directly, leaving their spatial and energy distributions largely unknown.
Therefore, existing models of Ganymede's ionosphere and exosphere require substantial assumptions or use measurements that cannot be unambiguously mapped to the regions of ionization.
This results in uncertainties in the ionization frequency of several orders of magnitude ($5\times10^{-8}-2\times10^{-5}\;\mathrm{s}^{-1}$, e.g. \cite{Eviatar2001,Marconi2007,Carnielli2019,CarberryMogan2023,Vorburger2024}).
For this reason, a sophisticated assessment of electron-impact ionization is necessary to investigate the spatial and compositional structure of Ganymede's ionosphere in detail.
In this study, we present a new method to constrain electron-impact ionization rates at Ganymede.
This method utilizes the relationship between electron-impact ionization and excitation rates of O$_2$.
Using observable auroral emissions in the ultraviolet, semi-forbidden OI 1356 \r{A} line, knowledge of electron fluxes or neutral oxygen abundances is not required.

Ganymede's aurora was discovered by HST observations of OI 1356 \r{A} and OI 1304 \r{A} emissions \citep{Hall1998,Feldman2000}.
As OI 1356 \r{A} doublet emissions are almost entirely caused by the excitation of O$_2$, the ratio to the OI 1304 \r{A} triplet emission was used to identify molecular oxygen as major constituent of Ganymede's atmosphere \citep{Hall1998,Molyneux2018} but also to identify other species, such as H$_2$O \citep{Roth2021}.
Initial observations supported the idea that the aurora primarily occurs at the polar caps, where Jovian particles have direct access to the atmosphere through the open magnetic field lines.
However, subsequent HST campaigns revealed that the aurora is structured into two bands near the two OCFB ovals of the magnetic environment \citep{McGrath2013}.
Contrary to the initial understanding, the polar caps were found to be relatively dark.
Statistical analysis of up to six HST campaigns carried out until 2017 with STIS showed temporal and spatial dependencies of the aurora brightness \citep{Musacchio2017,Marzok2022a}.
These results are subject to larger uncertainties considering the limited spatial resolution and the low signal-to-noise ratio.
Nevertheless, a spatial analysis of the temporal variation of the auroral ovals with Ganymede's position in relation to the Jovian plasma sheet could be used to remotely confirm the presence of a subsurface ocean on Ganymede \citep{Saur2015}.
Recently, the Juno UVS instrument provided substantially higher resolved images and confirmed the brightest emissions in bands spanning several degrees of latitude, which coincide with the locations of the OCFB \citep{Greathouse2022,Duling2022}.

This study is structured in the following sections.
In Section \ref{sec_ionization_excitation_ratio}, we begin with reviewing excitation and ionization cross sections of O$_2$.
Considering different electron energy distributions, we determine lower and upper limits for the ratio of OI 1356 \r{A} excitation to electron-impact ionization rates. 
From these we develop a new method to estimate ionization rates, applicable to any O$_2$ dominated atmosphere.
Section \ref{sec_ionization_rates_ganymede} presents an observational study of Ganymede's auroral OI 1356 \r{A} emissions during Juno's flyby and an application of the developed method to produce a global empirical map of ionization rates.
In Section \ref{sec_properties_ionosphere} we model density and outflow rates of Ganymede's ionosphere using the derived ionization rates and observed electron densities.
Section \ref{sec_discussion} concludes with a discussion of required ionizing electron populations and implications for the evolution of Ganymede's surface, as well as applied simplifications and limitations of our modeling.
In Section \ref{sec_summary} we summarize the main conclusions of this study.

\section{Ionization-to-Excitation Ratio}
\label{sec_ionization_excitation_ratio}

\subsection{Cross Sections}
OI 1356 \r{A} emissions result from the semi-forbidden $^3P \,\leftarrow \,^5S$ transition of atomic oxygen.
Oxygen in the $^5S$ state originates almost entirely from two different processes: (1) electron-impact dissociative excitation of O$_2$ and (2) electron-impact excitation of O.
The efficiency of both processes depends on the electron energy $W_e$.

\begin{figure}[htbp]
      \centering
      \includegraphics[width=\textwidth]{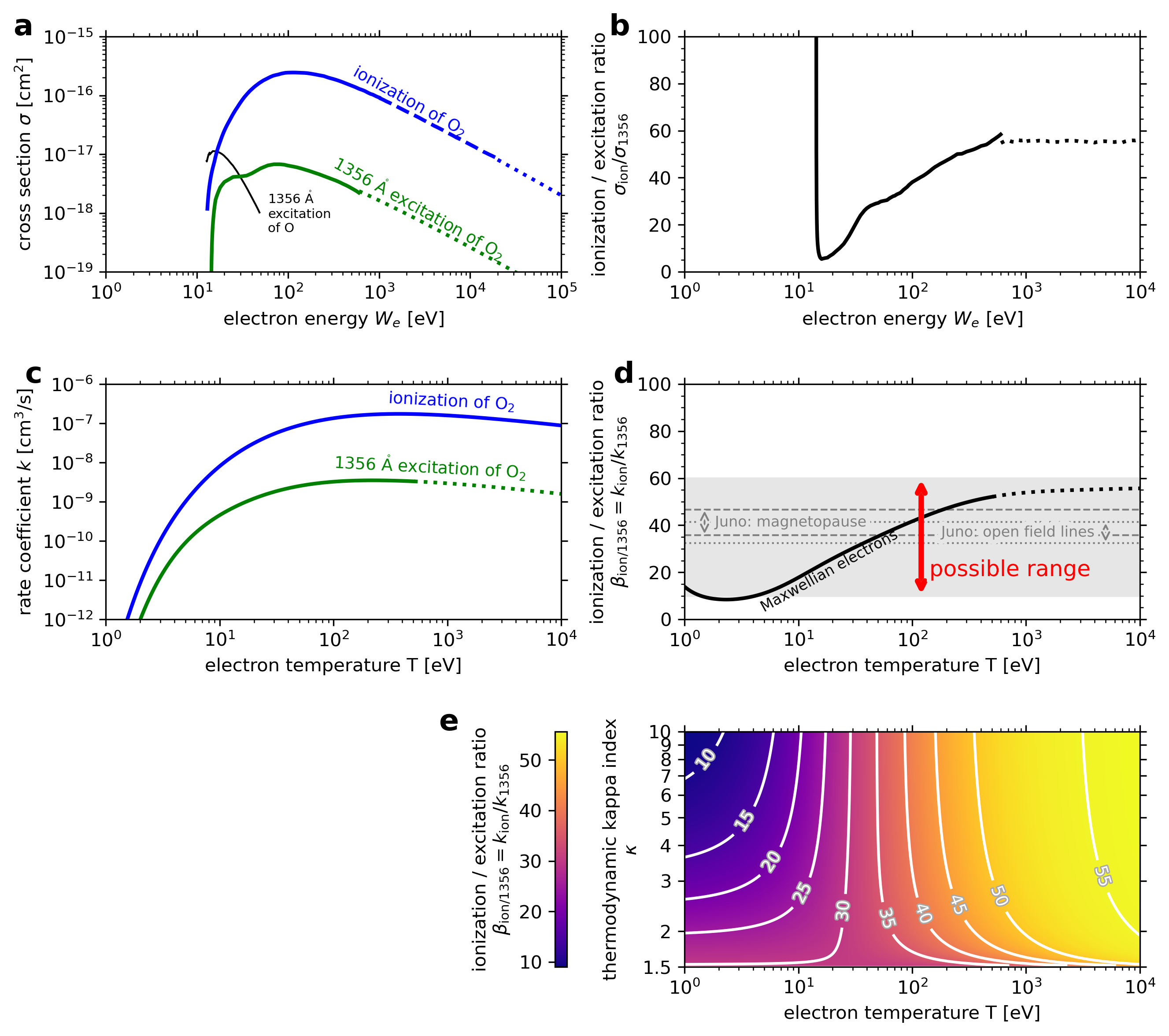}
      \caption{
      \textbf{a)} Cross sections for electron-impact ionization and OI 1356 \r{A} dissociative excitation of O$_2$ as function of electron energy from \cite{Lindsay2003} (blue), \cite{Schram1965} (dashed blue) and \cite{Kanik2003} (green).
      The measurements are extrapolated using Bethe-Oppenheimer relations (dotted).
      Excitation cross sections for atomic oxygen (black) are from \cite{Julienne1976}.
      \textbf{b)} Ratio of O$_2$ ionization and excitation cross sections.
      \textbf{c)} Interaction rate coefficients for Maxwell distributed electrons for ionization ($k_{\mathrm{ion}}$) and excitation ($k_{\mathrm{1356}}$), as function of temperature $T$.
      The dotted line indicates where $<$50\% of the electrons have energies with measured cross sections.
      \textbf{d)} Ratio of ionization and excitation rate coefficients, $\beta_{\mathrm{ion/1356}} = k_{\mathrm{ion}}/k_{\mathrm{1356}}$, for Maxwell distributed electrons as function of temperature $T$ (black).
      Gray lines indicate ranges for electron distributions measured by Juno close to the magnetopause and on open field lines, respectively \citep{Ebert2022a}.
      The ionization-to-excitation ratio is limited between 10 and 60 for all temperatures and distributions (gray shading).
      \textbf{e)} Ratio of ionization and excitation rate coefficients, $\beta_{\mathrm{ion/1356}} = k_{\mathrm{ion}}/k_{\mathrm{1356}}$, for Kappa distributed electrons as function of temperature $T$ and thermodynamic kappa index.
      }
      \label{fig_cross_section_ratio}
\end{figure}

The threshold energy for dissociative excitation of molecular oxygen,
\begin{equation}
      \mathrm{O}_2 + \mathrm{e}^-\rightarrow \mathrm{O}(^5S)+\mathrm{O}(^3P) + \mathrm{e}^-,
\end{equation}
is 14.3 eV \citep{Makarov2003}.
Electrons with higher energies can excite to higher states that cascade into $\mathrm{O}(^5S)$ or ionize the second $\mathrm{O}$. The threshold energy for 
\begin{equation}
      \mathrm{O}_2 + \mathrm{e}^-\, \rightarrow \,\mathrm{O}(^5S)+\mathrm{O}^+(^4S) + \mathrm{e}^-+\mathrm{e}^-
\end{equation}
is 27.9 eV \citep{Makarov2003}.
Total cross sections for the production of $\mathrm{O}(^5S)$ from all dissociative excitation processes of O$_2$ have been experimentally determined in the energy range 14-600 eV with an estimated error of 23\% (\cite{Kanik2003}, Figure \ref{fig_cross_section_ratio}a, green).
For higher energies we extrapolate the $>$100 eV data with a Bethe-Oppenheimer relation $AW_e^{-1}\ln\left(BW_e\right)$ (Figure \ref{fig_cross_section_ratio}a, dotted), obtaining $A=4.28\times10^{-16}\;\mathrm{cm}^2\mathrm{eV}$, $B=0.0474\;\mathrm{eV}^{-1}$.
Dissociative excitation of O$_2$ is most efficient from 20-200 eV with a maximum cross section of $6.8\times10^{-18}$ $\mathrm{cm}^2$ at around 80 eV.

Electron impact on atomic oxygen can either directly excite to the $\mathrm{O}(^5S)$ state ($\sim$25\%, \cite{Julienne1976}) or to higher states that cascade down to the $\mathrm{O}(^5S)$ state ($\sim$75\%).
Experimental cross sections for direct excitation are available in the 13.9-30 eV range \citep{Doering1989}.
Within their estimated uncertainty of $\pm$40\%, these measurements agree with theoretical predictions \citep{Julienne1976,Jackman1977}.
Total cross sections have been measured in the 10-80 eV range by \cite{Stone1974} and were later corrected by a factor of 2.8 \citep{Zipf1985}, resulting in a maximum of $9\times10^{-18}$ $\mathrm{cm}^2$.
Since these values are in reasonable agreement with the theoretical calculations by \cite{Julienne1976}, we adopt their total cross sections for our analysis, as shown in Figure \ref{fig_cross_section_ratio}a (black).
Theoretically, the threshold energy for exciting atomic oxygen is 9.14 eV, the energy of an emitted 1356 \r{A} photon.
Excitation of atomic oxygen is most efficient at around 15 eV ($11\times10^{-18}$ $\mathrm{cm}^2$) and decreases rapidly for higher energies.

To analyze the contribution of atomic oxygen to the OI 1356 \r{A} excitation at Ganymede, we must first acknowledge that it is only more effective than the excitation of molecular oxygen below 30 eV, rapidly declining for higher energies.
The total abundance of atomic oxygen in Ganymede's atmosphere is less than 10\% of the molecular oxygen density \citep{Hall1998,Roth2021}.
For equal excitation rates, the cross section for atomic oxygen must be ten times larger than that for O$_2$. 
This is only the case within an energy range of 5 eV between the threshold and 15.5 eV.
However, narrow electron distributions within this range are unlikely.
Furthermore, the OI 1356 \r{A} to OI 1304 \r{A} ratio does not indicate significant contributions of atomic oxygen \citep{Hall1998,Feldman2000,Molyneux2018,Roth2021}.
Similar to other studies estimating the contribution of atomic oxygen to be on the order of 1-3\% \citep{Tripathi2017}, we are confident that we can approximate that the OI 1356 \r{A} emissions at Ganymede are entirely from molecular oxygen.
Note, that in the cold and tenuous atmosphere of Ganymede each $\mathrm{O}(^5S)$ transitions back to $\mathrm{O}(^3P)$, so that we can equate $\mathrm{O}(^5S)$ excitation and OI 1356 \r{A} emission rates.

The threshold energy for electron-impact ionization of O$_2$ is 12.07 eV \citep{Tonkyn1989}.
The most recent measurements of cross sections in the 13-1000 eV energy range were carried out by \cite{Straub1996} with an estimated error of $\pm$3\%.
Cross sections from previous studies have a larger uncertainty of $\pm$13\% \citep{Kanik1993,Krishnakumar1992}.
\cite{Straub1996} measured the cross sections of three different ionization processes: the production of (1) O$_2^+$ ions, (2) O$^+$ and O$_2^{++}$ ions, and (3) O$^{++}$ ions.
The sum of these cross sections is interpreted as the total ionization cross section, with approximate, energy-dependent ratios of 64\%, 35\% and 1\%.
To account for a recalibration of the experimental apparatus, \cite{Lindsay2003} published a slightly modified dataset of the cross sections from \cite{Straub1996}, which we use in our analysis.
Figure \ref{fig_cross_section_ratio}a shows the total electron-impact ionization cross section as function of electron energy (blue).
For energies greater than 1000 eV, we use measurements in the 600-20,000 eV range \citep[][dashed]{Schram1965}.
Without modification, this data smoothly extends the 13-1000 eV data from \cite{Lindsay2003}.
For higher energies, we extrapolate the data from \cite{Schram1965} using a Bethe-Oppenheimer relation (dotted), obtaining $A=2.37\times10^{-14}\;\mathrm{cm}^2\mathrm{eV}$, $B=0.0448\;\mathrm{eV}^{-1}$.
Electron-impact ionization of O$_2$ is most efficient from 50 to 500 eV, reaching a maximum cross section of $2.5\times10^{-16}$ $\mathrm{cm}^2$ at approximately 120 eV.

\subsection{Ionization Rates from OI 1356 \r{A} Emissions}
\label{sec_ionization_rates_from_emissions}
Figure \ref{fig_cross_section_ratio}b shows the cross section ratio of electron-impact ionization and dissociative excitation of O$_2$.
Within the 14.4-600 eV energy range, this ratio is limited from 6 to 60.
Extrapolated cross sections introduce uncertainties for higher energies.
However, due to the declining magnitude of the cross sections, electrons with energies greater than 600 eV contribute minimally to interactions at Ganymede.
Below 14.4 eV, the ratio diverges, and only ionization is possible between 12.07 and 14.3 eV.
Because interactions within this small energy range are ineffective, these electrons do not significantly impact the total ionization-to-excitation ratio at Ganymede.

The energy distributions of the precipitating electrons that cause Ganymede's auroral emissions are unknown.
They are expected to vary both locally and temporally and are most likely not monoenergetic.
Therefore, we analyze the ratio of ionization to excitation rates by calculating the interaction rate coefficient,
\begin{equation}
      k=\int_{0}^{10^5\mathrm{ eV}} f(v)\sigma(v)v \,\mathrm{d}v.
      \label{eq_rate_coefficient}
\end{equation}
Here, $v=\sqrt{2W_e/m_e}$ is the electron velocity, $m_e$ is the electron mass and $\sigma$ is the cross section of the ionization or excitation interaction.
The isotropic speed distribution function $f$ is normalized such that $\int_{0}^{10^5\mathrm{ eV}} f(v) \,\mathrm{d}v=1$.
To minimize relativistic effects, we truncate the integration at 10$^5$ eV, where the difference between relativistic and non-relativistic velocities is $\sim$14\%.
For temperatures below 10$^4$ eV, the contribution of relativistic electrons is negligible, so this truncation does not affect the integration results.
Figure \ref{fig_cross_section_ratio}c shows the resulting ionization and excitation rate coefficients, $k_{\mathrm{ion}}$ and $k_{\mathrm{1356}}$, respectively, as function of Maxwellian temperature $T$.
Dotted lines indicate temperatures at which less than 50\% of the electrons have energies with measured cross sections.

For arbitrary electron distributions we can define
\begin{equation}
      \beta_{\mathrm{ion/1356}} = \frac{P_{\mathrm{ion}}}{P_{\mathrm{1356}}}=\frac{\nu_{\mathrm{ion}}n_{\mathrm{O}_2}}{\nu_{\mathrm{1356}}n_{\mathrm{O}_2}}=\frac{k_{\mathrm{ion}}n_en_{\mathrm{O}_2}}{k_{\mathrm{1356}}n_en_{\mathrm{O}_2}}=\frac{k_{\mathrm{ion}}}{k_{\mathrm{1356}}}
\end{equation}
to describe the ratio of electron-impact ionization and dissociative excitation rates $P$, where $\nu_{\mathrm{ion}}$ and $\nu_{\mathrm{1356}}$ are the ionization and excitation frequencies, respectively, and $n_e$ and $n_{\mathrm{O}_2}$ are the number densities of precipitating electrons and O$_2$.
Figure \ref{fig_cross_section_ratio}d (black line) shows $\beta_{\mathrm{ion/1356}}$ for the Maxwellian rate coefficients from Figure \ref{fig_cross_section_ratio}c as a function of electron temperature $T$.
Even over several orders of magnitude in temperature, the ratio increases from $\sim$10 to $\sim$60 only.
For temperatures $>$600 eV, the ratio has some uncertainties due to the lack of excitation cross section measurements; however, the available data indicates a flattening curve.
Only for temperatures $<$12 eV the ratio is smaller than 20.
At such low temperatures, a majority of the electrons are below the interaction thresholds, resulting in strongly reduced rate coefficients.
A Maxwell distribution with a ratio of 10 ($T<4$ eV) requires an electron density that is 13 times higher than that of a distribution with a ratio of 20 ($T\approx12$ eV) to excite the same emission brightness.
Therefore, one could argue that the ratio $\beta_{\mathrm{ion/1356}}$ is realistically confined between 20 and 60.

Measurements of Jovian electrons at Ganymede's orbit indicate non-Maxwellian distributions, often well described by a sum of two Kappa distributions for cold and hot electrons \citep{Pelcener2024,Sarkango2025}.
Figure \ref{fig_cross_section_ratio}e shows the ionization-to-excitation ratio $\beta_{\mathrm{ion/1356}}$ calculated by integrating a Kappa distribution as function of temperature and thermodynamic kappa index with Equation \ref{eq_rate_coefficient}.
For high values of $\kappa$ a Kappa distribution resembles a Maxwell distribution, confining $\beta_{\mathrm{ion/1356}}$ to the same limits of 10 to 60.
For $\kappa$ close to its lower limit of 3/2, where the contribution of the high energy tail is maximized, and the entropy is at its minimum, $\beta_{\mathrm{ion/1356}}$ is limited to a smaller range of 30 to 50.
Measurements upstream of Ganymede indicate a thermodynamic kappa index of 1.5$_{-0.0}^{+0.9}$ for the cold and 1.5$_{-0.0}^{+2.0}$ for the hot populations \citep{Sarkango2025}.
Considering that $\sim$90\% of the electrons are cold with a temperature of $\sim$137 eV, the Jovian electrons upstream of Ganymede are characterized with an ionization-to-excitation ratio of $\sim$34.

However, Ganymede is shielded from the Jovian plasma by its intrinsic magnetic field.
The origin of precipitating electrons causing the auroral emissions is uncertain and direct measurements are rare.
We examine distributions detected within Ganymede's magnetosphere by the Juno
Auroral Distribution Experiment (JADE) \citep{Ebert2022a}.
We consider field aligned electrons with pitch angle $<$45° or $>$135° detected in two regions:
(1) Electrons close to the magnetopause, which may be connected to the northern auroral oval,
(2) Electrons on open field lines, which probably precipitate up to 15° latitude north of the oval \citep{Duling2022}.
Because the maximums of the electron intensities are below the detection threshold of 32 eV, we linearly extrapolate the distributions in the low energy regime.
We consider three different assumptions for extrapolation: constant intensities, decreasing intensities (zero at 10 eV) and increasing intensities (at the magnetopause 10$^{10}$ and on open field lines 10$^{9}$ at 10 eV, in units of (cm$^{2}$ sr s keV)$^{-1}$).
Because cross sections for these low energies are small, calculated interaction frequencies would depend significantly on the extrapolation.
However, the extrapolations only have a negligible impact on the ionization-to-excitation ratio $\beta_{\mathrm{ion/1356}}$ ($<$10\%).
For all distributions available from \cite{Ebert2022a} and all different extrapolations, we calculate the extreme values to be $36<\beta_{\mathrm{ion/1356}}<47$ (magnetopause) and $32<\beta_{\mathrm{ion/1356}}<41$ (open field lines), respectively (see the dashed and dotted ranges in Figure \ref{fig_cross_section_ratio}d).

Hence, for realistic electron distributions, the ratio of electron-impact ionization of O$_2$ and OI 1356 \r{A} excitation frequencies (or rates) at Ganymede is always limited between approximately 10 and 60.
This means, if the excitation frequency is known, then the ionization frequency is constrained with an uncertainty factor of only 6:
\begin{equation}
      \nu_{\mathrm{ion}} = \beta_{\mathrm{ion/1356}} \;\nu_{\mathrm{1356}}.
\end{equation}
In other words, for each 1356 \r{A} photon emitted, there are $\beta_{\mathrm{ion/1356}}$ (10-60) ionization interactions.

Available observations of auroral emissions at Ganymede do not directly provide excitation frequencies.
Instead, the emission brightness,
\begin{equation}
      B = k_{\mathrm{1356}} n_e N_{\mathrm{O}_2},
      \label{eq_definition_brightness}
\end{equation}
is the natively measured quantity, given in units of $\mathrm{R}=10^{10}\mathrm{photons}/\mathrm{m}^2\mathrm{s}$ and interpreted as number of emitted photons within the column of line of sight of the instrument.
The brightness can be transformed for a column perpendicular to Ganymede's surface.
For the remainder of this article, we will use this orientation, so that $N_{\mathrm{O}_2}$ is the radial column density of Ganymede's O$_2$ atmosphere.
Similarly to the frequencies, we can utilize the known ratio $\beta_{\mathrm{ion/1356}}$ to calculate the column rate of electron-impact ionization processes of O$_2$, or column production rate,
\begin{equation}
      \Pi_{\mathrm{ion}}=k_{\mathrm{ion}} n_e N_{\mathrm{O}_2},
      \label{eq_definition_column_production_rate}
\end{equation}
from brightness measurements of OI 1356 \r{A} emissions with
\begin{equation}
      \Pi_{\mathrm{ion}} = \beta_{\mathrm{ion/1356}} B.
      \label{eq_ionization_from_brightness}
\end{equation}
With this approach, the uncertainty in calculating electron-impact ionization rates, usually arising from estimations of ionization frequency and atmospheric densities, is reduced to a factor of less than 6.

\subsection{Uncertainty of the Ionization-to-Excitation Ratio}
Because the ionization-to-excitation cross section ratio is already limited to the range of 10 to 60 for all energies, and interaction rates are calculated with the integration of Equation \ref{eq_rate_coefficient}, the ionization-to-excitation ratio $\beta_{\mathrm{ion/1356}}$ must always be confined between 10 and 60 for any electron energy distribution.
The only exception to this are nearly monoenergetic electrons within the small energy range between the ionization and excitation thresholds ($\sim$5 eV width), where ionization is very ineffective (Figure \ref{fig_cross_section_ratio}b).

Nevertheless, a qualitative assessment of the uncertainty in estimating the ratio of a whole atmospheric column from a single electron distribution can be made.
Considering column interaction rates, Equations \ref{eq_definition_brightness} and \ref{eq_definition_column_production_rate} assume that electron distributions are independent of altitude $h$.
Before we discuss the impact of interactions at higher altitudes on the ratio at lower altitudes, we first assess the attenuation of the precipitating electron flux through Ganymede's atmosphere.
To do so, we calculate the interaction depth ("electron optical depth") of Ganymede's atmosphere with regard to a certain type of interactions with O$_2$,
\begin{equation}
      \tau = \int_0^\infty n_{\mathrm{O}_2}(h) \sigma(W_e(h)) \mathrm{d}h \approx N_{\mathrm{O}_2} \sigma,
\end{equation}
where $\sigma$ is the corresponding cross section.
Using upper limits of $N_{\mathrm{O}_2\mathrm{,max}}=10^{15}$ cm$^{-2}$ and $\sigma_{\mathrm{ion,max}}=2.5\times10^{-16}$ $\mathrm{cm}^2$ (at 120 eV), we estimate a maximum possible interaction depth of $\tau_{\mathrm{ion,max}}=0.25$ for ionization.
This results in a transmission of $F/F_0 = e^{-\tau_{\mathrm{ion,max}}}=78\%$, where $F$ and $F_0$ are the electron fluxes at the surface and at the top of the atmosphere, respectively.
This means that at least 78\% of the precipitating electrons reach the surface without an ionization interaction.
Taking momentum-transfer collisions into account as well, where the maximum cross section is $\sigma_{\mathrm{col,max}}=6.8\times10^{-16}$ $\mathrm{cm}^2$ (at 1.5 eV) \citep{Elford2003}, the transmission reduces to 51\%.
Regarding momentum-transfer collisions, the transmission increases rapidly with higher energies, as these interactions are more effective than ionization only at energies $<$~70 eV.
Considering that the estimated transmission rates are lower limit values, we can confidently consider Ganymede's atmosphere to be thin with respect to interactions with precipitating electrons.

We now qualitatively discuss how the energy distribution and ionization-to-excitation ratio vary with altitude due to energy degradation, secondary electron cascades, and pitch-angle scattering.
Ionization, excitation, and momentum-transfer collisions reduce the energy of electrons, thereby altering the energy distribution at lower altitudes.
Ionization produces secondary electrons, which can further interact with neutral oxygen and cascade to even lower energies.
However, these secondary electrons typically receive only a small fraction of the remaining energy and are therefore relatively cold.
Because Ganymede's atmosphere is thin, these processes shift the energy distribution only slightly toward lower energies, where the ionization-to-excitation cross section ratio remains within the range of 10 to 60.
As a result, energy degradation may lower 
$\beta_{\mathrm{ion/1356}}$ at lower altitudes, but this effect is negligible and smaller than the uncertainty associated with estimating the initial energy distribution of the precipitating electrons.
First-order modeling indicates that using a constant $\beta_{\mathrm{ion/1356}}$ for the entire atmospheric column, based on Maxwellian electrons at the top of the atmosphere, introduces a maximum error of 2.5\% compared to a self-consistent calculation with altitude-dependent energy distributions (see Supporting Information).
Although pitch angle scattering can cause electrons to remain longer in the atmosphere, it does not alter their energy and therefore has no direct effect on the ionization-to-excitation ratio.
Additionally, pitch angle diffusion coefficients on open field lines are very small \citep{Williams1997b} and on closed field lines they are unknown.

Therefore, we conclude that the ionization-to-excitation ratio  $\beta_{\mathrm{ion/1356}}$ consistently remains between 10 and 60, with energy degradation effects having a negligible impact on the ratio for an entire atmospheric column. 
This shows that the method presented in Section \ref{sec_ionization_rates_from_emissions} is a robust and powerful tool for estimating electron-impact ionization rates from OI 1356 \r{A} emission brightness observations at Ganymede.

\section{Ionization Rates at Ganymede}
\label{sec_ionization_rates_ganymede}

\subsection{OI 1356 \r{A} brightness observed by Juno}
\label{sec_ionization_rates_from_juno}

\begin{figure}[htbp]
      \centering
      \includegraphics[width=0.97\textwidth]{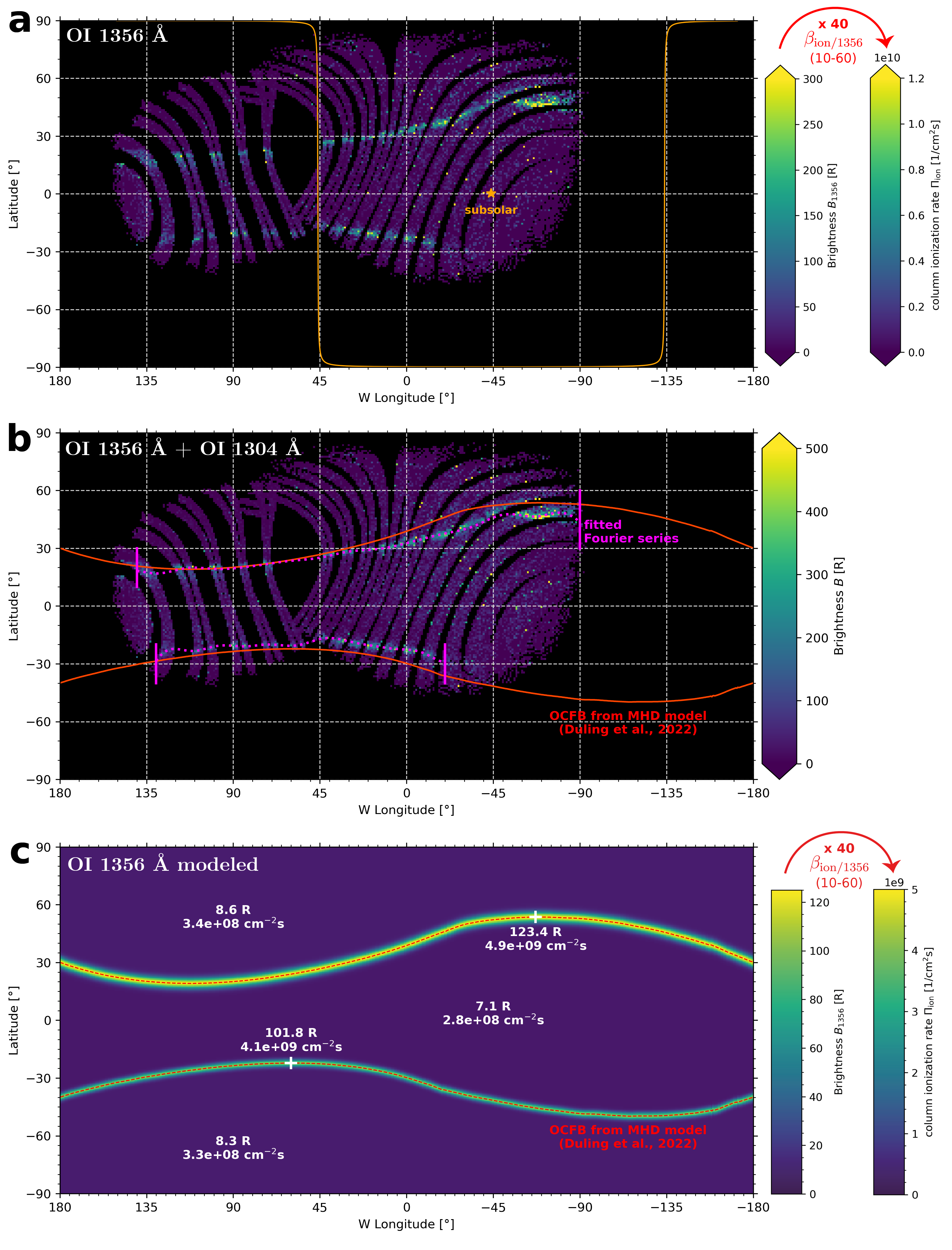}
      \caption{
      \textbf{a)}
      OI 1356 \r{A} emission brightness as observed by Juno UVS during the PJ34 flyby.
      Reflected sunlight has been subtracted.
      The color bar is saturated, 56 pixels have values above 300 R, 15 pixels above 500 R.
      The orange line shows the terminator.
      The right color bar shows electron-impact ionization rates of O$_2$ for an ionization-to-excitation ratio of $\beta_{\mathrm{ion/1356}}=40$, characterizing measured electron distributions.
      Possible values for the ratio are 10-60.
      \textbf{b)}
      Integrated OI 1356 \r{A} and OI 1304 \r{A} emission brightness, reflected sunlight has been subtracted.
      The color bar is saturated, 35 pixels have values above 500 R, 3 pixels above 1000 R.
      The red lines show the modeled location of the open-closed-field line-boundary (OCFB) \citep{Duling2022}.
      The vertical purple lines indicate the longitude range considered for fitting a Fourier series to the brightest emissions, shown as dotted purple lines.
      \textbf{c)}
      Empirically modeled OI 1356 \r{A} brightness and column electron-impact ionization rates of O$_2$.
      The model uses the OCFB from the MHD model of \cite{Duling2022} and average brightnesses from Figure a, calculated with the fitted OCFB from Figure b (Equation \ref{eq_gauss_fit}).
      White numbers show constant background values and peak values of the auroral ovals.
      }
      \label{fig_1356_observation_map}
\end{figure}

From Juno UVS brightness observations of Ganymede's 1356 \r{A} aurora we calculate O$_2$ ionization rates by applying the method described in Section \ref{sec_ionization_excitation_ratio}.
Juno UVS collects time-tagged pixel list data, with location, wavelength, and time information available for each photon detected \citep{Gladstone2014}.
We used this data to produce a OI 1356 \r{A} map (Figure \ref{fig_1356_observation_map}a), following the general method used by \cite{Greathouse2022} for their integrated OI 1304 \r{A} and OI 1356 \r{A} images.
As in the previous study, we selected photons measured through the two 2.55° x 0.2° wide portions of the “dog-bone”-shaped UVS slit only, ignoring the narrow 2.0° x 0.025° middle section where the signal-to-noise ratio is low.
\cite{Greathouse2022} applied a -35 ms shift to the Juno spin phase in order to correct a seemingly nonphysical kink in the southern aurora.
We used a reduced shift of -17 ms, as \cite{Waite2024} found that this value moves the aurora closer to the surface, where their models suggest the emissions should peak.
However, the new time shift does not fully remove the kink in the southern band near the terminator (shown in orange) and reduces the fit to the modeled OCFB location (see Figure \ref{fig_1356_observation_map}b).

On Ganymede's day side, a small fraction of the detected photons may be attributed to reflected sunlight.
To correct for this, we also produced a map of detected 1335 \r{A} photons, corresponding to the wavelength of a CII emission line in the solar spectrum.
This emission is not expected to be produced locally at Ganymede and has previously been used to scale and subtract reflected sunlight from Earth-based observations of Ganymede (e.g., \cite{Hall1998}; \cite{Roth2021}).
We used a TIMED/SEE spectrum \citep{Woods2005} to determine the solar 1335 \r{A} / 1356 \r{A} ratio during the Juno UVS observations.
We scaled our 1335 \r{A} image using this ratio (1335 \r{A} / 1356 \r{A} = 4.3), resulting in a map of the expected contribution of reflected sunlight at 1356 \r{A}, which we then subtracted from the OI 1356 \r{A} auroral image. 

To account for the low signal-to-noise ratio, we further removed pixels with exposure times smaller than 20 ms from the data.
In the final, sunlight-corrected map (Figure \ref{fig_1356_observation_map}a), the aurora appears as two sharp ovals with brightness values up to 500 R.
Low count rates still cause significant Poisson noise, potentially explaining individual pixels with up to 1100 R.

Multiplying the OI 1356 \r{A} brightness by $10<\beta_{\mathrm{ion/1356}}<60$ directly gives the column ionization rates of O$_2$ (Equation \ref{eq_ionization_from_brightness}).
The exact value of $\beta_{\mathrm{ion/1356}}$ depends on the electron distribution.
Spatial and temporal variations in the electron distributions are unknown but possible, particularly when comparing the regions of open and closed field lines with the bright auroral ovals.
Therefore, it can be expected that $\beta_{\mathrm{ion/1356}}$ varies within its limits across the map.
For demonstration purposes we use an intermediate value of $\beta_{\mathrm{ion/1356}}=40$, which we also calculated for measured distributions (see Section \ref{sec_ionization_rates_from_emissions}).
The resulting ionization rates are shown with the second color bar in Figure \ref{fig_1356_observation_map}a.

\subsection{Global Empirical Model for Ionization Rates}
\label{sec_ionization_model}

In Section \ref{sec_ionization_rates_from_juno} we calculated ionization rates directly from brightness values of the Juno UVS observations.
However, due to Poisson noise and the lack of global coverage, the resulting map is impractical for modeling applications.
Therefore, we use the OI 1356 \r{A} brightness data to derive an empirical global model for the emission and ionization rates at Ganymede.

Figure \ref{fig_1356_observation_map}b shows the modeled location of the OCFB on Ganymede's surface \citep{Duling2022} (red), in comparison with the integrated OI 1356 \r{A} and OI 1304 \r{A} brightness data, similar to Figure \ref{fig_1356_observation_map}a.
While the modeled OCFB aligns well with the bright ovals, modeling uncertainties result in notable discrepancies, particularly on the trailing side.
It is an open question whether the brightest emissions occur directly at the OCFB or if the OCFB is a poleward boundary of the ovals, as partly suggested by previous studies \citep{Greathouse2022}.
Here, we estimate the location of the OCFB during Juno's flyby by fitting two Fourier series in longitude $\lambda$ to the latitudes $\theta$ of the brightest emissions in the northern and southern hemispheres.
We use the body fixed \textit{IAU\_Ganymede} system, in which longitude $\lambda$ is measured positively west from the Jupiter-facing meridian.


The fitting procedure consists of two consecutive steps.
First, we manually pick multiple coordinates at the center of the ovals, which we then fit using a second-order Fourier series and a least squares algorithm.
In the second step, we then select all pixels within a 7° latitude distance of the initial fit that are brighter than 100 R and at longitudes well covered by the observation, i.e. $140$°$>\lambda>-90$° (north) and $130$°$>\lambda>-20$° (south).
Using the integrated OI 1356 \r{A} and OI 1304 \r{A} brightness as weights for the coordinates of the selected pixels, we fit a third-order Fourier series with a higher-order extension to both ovals:
\begin{equation}
      \theta_\mathrm{OCFB,fit}(\lambda)=\sum_{i=0}^{3}A_i\sin\left(i\lambda+\Delta\lambda_i\right)+\sum_{i=10}^{12}A_i\sin\left(i\lambda+\Delta\lambda_i\right).
      \label{eq_fourier_series}
\end{equation}
Due to uncertainties in the mapping of detected photons and the likely dynamics of the OCFB location during the 12-minute observation period (see Section \ref{sec_variability}), the brightest emissions do not apparently form smooth ovals.
To account for this and still match the brightest emissions even on small scales, the higher-order extension in Equation \ref{eq_fourier_series} is necessary.
The resulting fits for the ovals are shown with dotted purple lines in Figure \ref{fig_1356_observation_map}b for the longitude ranges considered in the analysis.
Note that we exclude pixels with $\lambda<-20$° in the southern hemisphere from the data because an emission band is barely visible there.

Apart from a bright spot of the northern oval at -70° W, the brightness of the ovals does not reveal a strong longitudinal dependence.
This contrasts with the findings of \cite{Marzok2022a}, whose statistical analysis of HST observations from 1998 to 2017 suggested that emissions at the sub- and anti-Jovian sides are four times fainter than those at the leading and trailing sides. 
This discrepancy may result from the limited number of exposures covering the sub- and anti-Jovian regions in the \cite{Marzok2022a} dataset - only 6 out of 46 total exposures - as well as the lower spatial resolution of those observations.
Therefore, aside from the location of the OCFB, we do not consider possible longitudinal variations in our analysis.

\begin{figure}[htbp]
      \centering
      \includegraphics[width=1.0\textwidth]{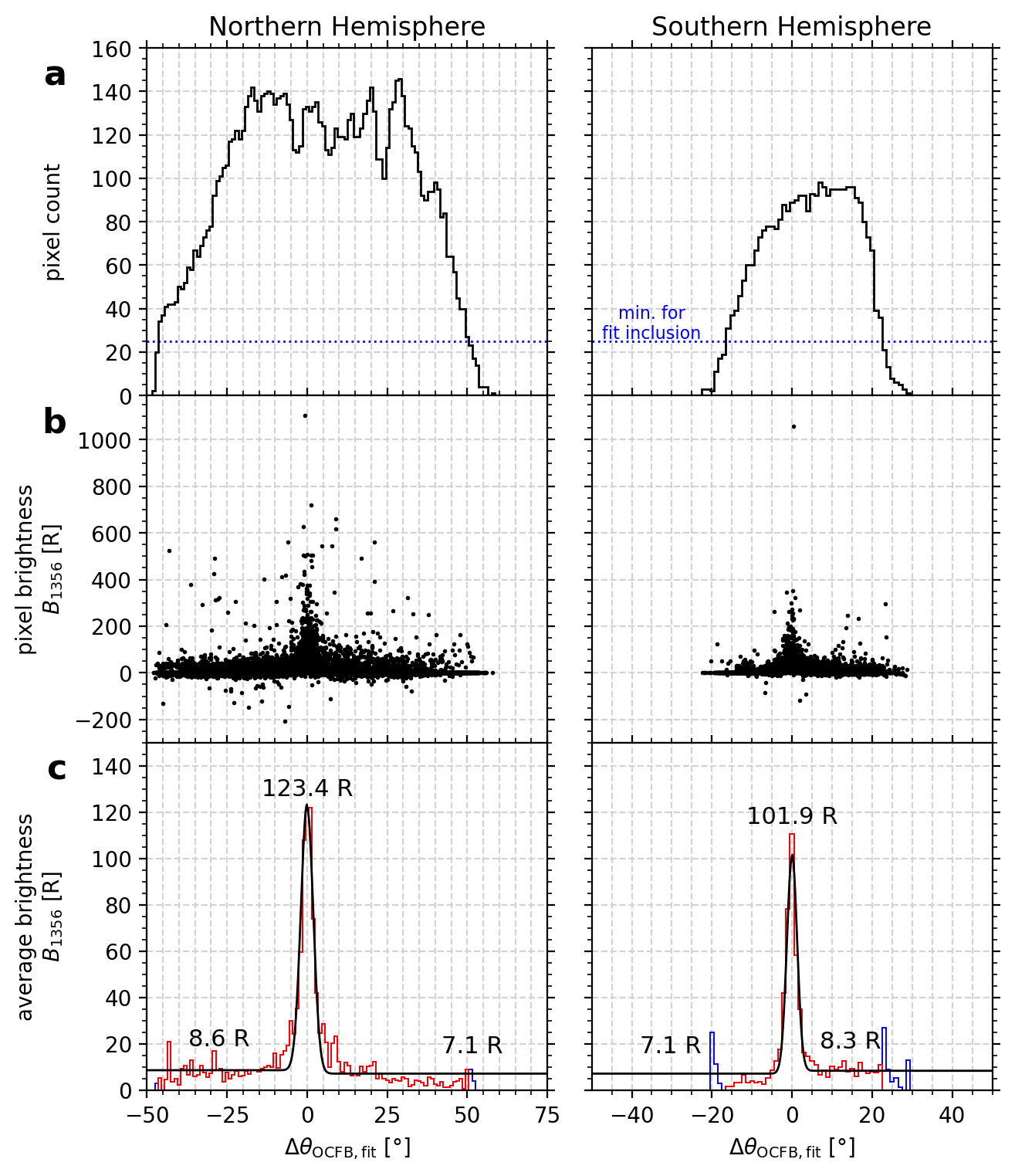}
      \caption{
      OI 1356 \r{A} brightness as function of latitudinal distance $\Delta\theta_\mathrm{OCFB,fit}$ from the fitted OCFB (Figure \ref{fig_1356_observation_map}b).
      The left side shows the northern hemisphere, the right side the southern hemisphere.
      \textbf{a)} Number of pixels in each 1° wide bin of latitude.
      \textbf{b)}: Brightness of each pixel sorted by 1° wide bins of latitude.
      \textbf{c)}: OI 1356 \r{A} emission brightness averaged over all longitudes as function of $\Delta\theta_\mathrm{OCFB,fit}$ (red, blue).
      Data of bins with $>$25 pixel counts (red) are fitted with Gaussian distributions at the OCFB and constants in the background region (black).
      The numbers show fitted background values and average peak emissions at the center of the auroral ovals.
      }
      \label{fig_uvs_data_latitudinal_dependence}
\end{figure}

We continue to analyze the morphology of the OI 1356 \r{A} emissions in the latitudinal direction, separately for the northern and southern hemispheres.
To accomplish this, we sort all pixels into 1° wide bins according to their latitudinal distance from the fitted OCFB, $\Delta\theta_\mathrm{OCFB,fit} = \theta_\mathrm{OCFB,fit}- \theta$, shown in Figure \ref{fig_uvs_data_latitudinal_dependence}b.
For pixels with a brightness of exactly 0 R, no 1356 \r{A} photons were detected, and no reflected sunlight was subtracted.
We calculate negative brightness values for pixels where non-zero count rates of 1335 \r{A} photons suggest higher rates of reflected 1356 \r{A} photons than are detected.
However, this only appears unphysical at first, as this statistical consequence of the Poisson noise correctly averages out over larger spatial scales.
Figure \ref{fig_uvs_data_latitudinal_dependence}a shows the number of available pixels in each bin.
We then calculate the average brightness of each bin to determine the OI 1356 \r{A} brightness averaged over all longitudes as function of $\Delta\theta_\mathrm{OCFB,fit}$, shown in \ref{fig_uvs_data_latitudinal_dependence}c in red and blue.

To obtain an analytic expression, we fit the average brightness values of Figure \ref{fig_uvs_data_latitudinal_dependence}c with three constants for the regions north, south and between the two ovals, called background region in the remainder of this article.
For the ovals we overlay the background with two Gaussians centered at the fitted OCFB.
To reduce the impact of noisy pixels with high values, we exclude bins with a pixel count smaller than 25 (blue in Figure \ref{fig_uvs_data_latitudinal_dependence}c), so that only the red data is used for the fitting.
The resulting average brightness as a function of $\Delta\theta_\mathrm{OCFB,fit}$ is shown by the black line in Figure \ref{fig_uvs_data_latitudinal_dependence}c.
Given the locations of the northern and southern OCFB, $\theta_\mathrm{OCFB,N}$ and $\theta_\mathrm{OCFB,S}$, respectively, the average brightness as a function of longitude $\lambda$ and latitude $\theta$ can then be modeled as
\begin{equation}
      B_\mathrm{1356}\left(\theta, \lambda\right) = 
      \left\{\begin{array}{ll}
        8.6 \mathrm{\;R}, & \mathrm{for\;} \theta>\theta_\mathrm{\scriptscriptstyle OCFB,N}(\lambda)\\
        7.1 \mathrm{\;R}, & \mathrm{for\;} \theta_\mathrm{\scriptscriptstyle OCFB,N}(\lambda)\ge\theta\ge \theta_\mathrm{\scriptscriptstyle OCFB,S}(\lambda)\\
        8.3 \mathrm{\;R}, & \mathrm{for\;} \theta< \theta_\mathrm{\scriptscriptstyle OCFB,S}(\lambda)
        \end{array}\right\} +
        \left\{\begin{array}{ll}
        114.8 \mathrm{\;R} \exp\left(\frac{-\left(\theta_\mathrm{OCFB,N}(\lambda)-\theta\right)^2}{2\left(1.94\degree\right)^2}\right), & \mathrm{for\;} \theta>0\degree \\
        93.5 \mathrm{\;R} \exp\left(\frac{-\left(\theta_\mathrm{OCFB,S}(\lambda)-\theta\right)^2}{2\left(1.28\degree\right)^2}\right), & \mathrm{for\;} \theta<0\degree
        \end{array}\right\}.
        \label{eq_gauss_fit}
\end{equation}

The peak emissions centered at the fitted OCFB clearly exhibit a Gaussian shape, with widths (FWHW) of 4.6° (north) and 3.0° (south).
However, the actual location of the OCFB during the observations is unknown.
Therefore, it is possible that the Gaussian distribution is caused by statistical deviations in estimating the OCFB location.
Nevertheless, the emissions appear relatively symmetric around the brightest location.
The ovals do not show a sharp poleward decay and gentle decline toward the equator, as indicated on the leading side by the analysis of \cite{Greathouse2022}, although our analysis cannot completely rule this out due to the averaging over all longitudes.
In the northern hemisphere, at a distance of 5-10° from the OCFB, the distribution is wider than our Gaussian fit.
This may be due to more dispersed emissions on the trailing side, indicating either a broader flux of precipitating electrons or stronger dynamics of its location.
The corresponding region of the southern oval was not observed, which may explain the better fit to the Gaussian distribution there.
However, the characteristics of the emissions near the southern oval on the trailing side remain unknown.

In the background region poleward of the OCFB, the data show relatively homogeneous brightness values, which are consistent with our constant fit.
In these regions of open field lines, the average background brightness is $\sim$8.5 R, which is consistent with the results of \cite{Waite2024}.
However, in the equatorial region the brightness declines towards the equator, a trend that is not captured by our model.
With an average brightness of $\sim$7 R, the equatorial region, which is exposed to closed field lines, is slightly darker.
On average, the northern auroral oval peaks at $\sim$125 R, and the southern oval peaks at $\sim$100 R.
This asymmetry is consistent with the observation that auroral emissions are brighter in the hemisphere closer to Jupiter's current sheet \citep{Saur2022}.


Figure \ref{fig_1356_observation_map}c shows the global empirical model of OI 1356 \r{A} emissions based on Equation \ref{eq_gauss_fit} and the OCFB location from \cite{Duling2022}, which is suited for the situation during Juno's flyby.
Apart from the variation of the OCFB location, our analysis does not investigate brightness variations with longitude.
Due to lack of observational coverage, the modeled morphology of Equation \ref{eq_gauss_fit} is mostly based on the leading and sub-Jovian sides.
However, the aim of this study is not to analyze the morphology of Ganymede's auroral emissions in detail.
Rather, Equation \ref{eq_gauss_fit} can be used to model average electron-impact ionization rates on a global scale.
To calculate column ionization rates, the brightness is multiplied by an ionization-to-excitation ratio $\beta_{\mathrm{ion/1356}}$ between 10 and 60, as described in Section \ref{sec_ionization_excitation_ratio}.
The right color bar in Figure \ref{fig_1356_observation_map}c shows ionization rates in case an intermediate value of $\beta_{\mathrm{ion/1356}}=40$ is used, i.e. similar electron distributions are assumed across the entire map.
The resulting rates are $\sim$$3\times10^{8}$ cm$^{-2}$s$^{-1}$ in the background regions and up to $\sim$$5\times10^{9}$ cm$^{-2}$s$^{-1}$ within the bright auroral bands.

Our model features a global OI 1356 \r{A} emission rate of $\sim$$1.3\times10^{25}\mathrm{s}^{-1}$.
We validate this value using photon fluxes observed by \cite{Feldman2000} with the HST STIS instrument.
During several exposures, $F=15-40\times10^{-5}$ photons/cm$^2$s of the OI 1356 \r{A} doublet reached the detector.
However, since the direction of emitted photons is distributed equally, approximately every second photon interacts with Ganymede's surface.
The albedo of icy moon surfaces in the far ultraviolet is very small, $\sim$0.015 \citep{Hall1998, Saur2011}, so that these photons are absorbed and not detectable.
This increases the total number of emitted photons by a factor of two.
Assuming isotropic radiation and an optically thin atmosphere, the global OI 1356 \r{A} emission rate is estimated by integrating the detected photon flux $F$ over the surface of a sphere spanned by the distance between Ganymede and HST during observation ($d\approx4.25 \;\mathrm{AU}$).
The resulting global emission rate is $8\pi d^2F \approx 1.5-4.0\times 10^{25}  \;\mathrm{s}^{-1}$, which is slightly higher but on the same order as the emission rate of our model.
The corresponding OI 1356 \r{A} luminosity is $1.0-2.9\times10^{7}\;\mathrm{W}$, consistent with $3\times10^{7}\;\mathrm{W}$ for OI 1356 \r{A} and OI 1304 \r{A} emissions calculated by \cite{Saur2021a}.

The global electron-impact ionization rate of O$_2$ in our model is $\sim$$5\times10^{26}\;\mathrm{s}^{-1}$, with lower and upper limits of $1.3-7.6\times10^{26}\;\mathrm{s}^{-1}$ for extreme values of $\beta_{\mathrm{ion/1356}}$.
Based on the HST photon flux, we estimate absolute limits of $1.5-24.0\times10^{26}\;\mathrm{s}^{-1}$.
In our model, the auroral ovals (49 \%, defined by a width of three standard deviations) and the background regions (51\%) contribute nearly equally to the global ionization rate, while the ovals cover only 13\% of Ganymede's surface.

\section{Properties of Ganymede's Ionosphere}
\label{sec_properties_ionosphere}
We use the ionization rates calculated in the previous section to evaluate the ionospheric density at Ganymede and the role of transport.
As a reference case, we first neglect transport and calculate densities assuming chemical equilibrium between ionization and recombination.
We then assess the role of transport by comparing modeled densities with available measurements.

\subsection{Chemical Equilibrium Versus Transport}
\label{sec_density_model}

We neglect the ionization of species other than O$_2$, because they are minor constituents of Ganymede's atmosphere, or, in the case of H$_2$O, are confined to the subsolar point.
Thus, in this study, we consider a pure O$_2^+$ plasma as lower limit of the ionospheric density and neglect other potentially significant constituents \citep[e.g.][]{Beth2025}.
By treating electrons and O$_2^+$ ions as a plasma, we model its number density $n$ with a continuity equation of the form
\begin{equation}
      \partial_t n + \nabla \cdot (n \mathbf{v}) = P - L,
      \label{eq_continuity}
\end{equation}
where $P$ and $L$ are volumetric production and loss rates, respectively, and $\mathbf{v}$ is the bulk velocity.
Since we are considering a quasi-neutral plasma consisting only of O$_2^+$ and electrons, $n$ denotes the density of both the ionospheric electrons and ions.
The density of precipitating electrons is described by $n_e$.
For simplicity, we assume stationarity and neglect the transport $n\mathbf{v}$.
We then calculate the chemical equilibrium between ionization and recombination:
\begin{equation}
      P = L .
\end{equation}
Both electron-impact and photoionization contribute to the plasma production rate,
\begin{equation}
      P = P_\mathrm{ion} + P_{\mathrm{h\nu}}.
\end{equation}

We adopt an O$_2$ atmosphere in hydrostatic equilibrium with a spatially constant scale height of $H=20 \;\mathrm{km}$.
Then, we use the column ionization rates from Equation \ref{eq_ionization_from_brightness} to calculate the volumetric ionization rates as a function of altitude $h$:
\begin{equation}
      P_\mathrm{ion}\left(h, \theta, \lambda \right) = 
      \frac{\Pi_\mathrm{ion}\left(\theta, \lambda \right)}{H}\mathrm{e}^{-h/H}=
      \frac{\beta_{\mathrm{ion/1356}}\left(\theta, \lambda \right)B_\mathrm{1356}\left(\theta, \lambda \right)}{H}\mathrm{e}^{-h/H}.
\end{equation}

The photoionization rate is well constrained with a frequency of $\nu_{\mathrm{h\nu}}\approx3 \times 10^{-8} \;\mathrm{s}^{-1}$ \citep{Huebner1992},
\begin{equation}
      P_{\mathrm{h\nu}}\left(h, \theta, \lambda \right) 
      = \nu_{\mathrm{h\nu}} \frac{N_{\mathrm{O}_2}}{H} \mathrm{e}^{-h/H}S\left(h, \theta, \lambda \right),
\end{equation}
where $S=1$ in the illuminated, optically thin atmosphere and $S=0$ on the night side.
Unlike our method of quantifying electron-impact ionization, knowledge of the atmospheric column density $N_{\mathrm{O}_2}$ is necessary to quantify photoionization.
Here, we adopt $N_{\mathrm{O}_2}=5\times10^{14}\;\mathrm{cm}^{-2}$, consistent with recent modeling and observations \citep{Kleer2023,Leblanc2023,Vorburger2024,Milby2024}.
Even for the upper limit of $10^{15}\;\mathrm{cm}^{-2}$ the column ionization rate from photoionization is only $\sim3\times10^{7}$ cm$^{-2}$s$^{-1}$, a factor of $>$10 smaller than the low electron-impact ionization rate associated with the faint OI 1356 \r{A} emission in the background region (8 R).

The loss rate from dissociative recombination is calculated as
\begin{equation}
      L(h,\theta,\lambda) = \alpha \left[n(h,\theta, \lambda)\right]^2.
\end{equation}
Since low-energy electrons recombine more efficiently, the recombination coefficient $\alpha$ depends on the electron temperature $T$.
Here, we adopt an estimate of $\sim$1000 K, which leads to $\alpha=10^{-7} \;\mathrm{cm}^3 \mathrm{s}^{-1}$, which is about 20\% of its maximum possible value resulting from 100 K electrons \citep{Sheehan2004}.

The resulting chemical equilibrium density of O$_2^+$ is given by:
\begin{equation}
      n_{\mathrm{chem}} = 
      n_{\mathrm{chem,0}} \mathrm{e}^{-h/2H}.
\label{eq_plasma_density}
\end{equation}
It falls off with the double scale height of the atmosphere and has a surface value of
\begin{equation}
      n_{\mathrm{chem,0}} =
                  \sqrt{
                  \frac{\beta_{\mathrm{ion/1356}}B_\mathrm{1356}}{\alpha H }
                 +\frac{\nu_{\mathrm{h\nu}}N_{\mathrm{O}_2}}{\alpha H}S.
      }
      \label{eq_surface_density}
\end{equation}
Notably, with the method of this article the equilibrium density of O$_2^+$ can be estimated independently of the neutral column density when ionization is dominated by electron-impact interactions.

\begin{figure}[htbp]
      \centering
      \includegraphics[width=1.0\textwidth]{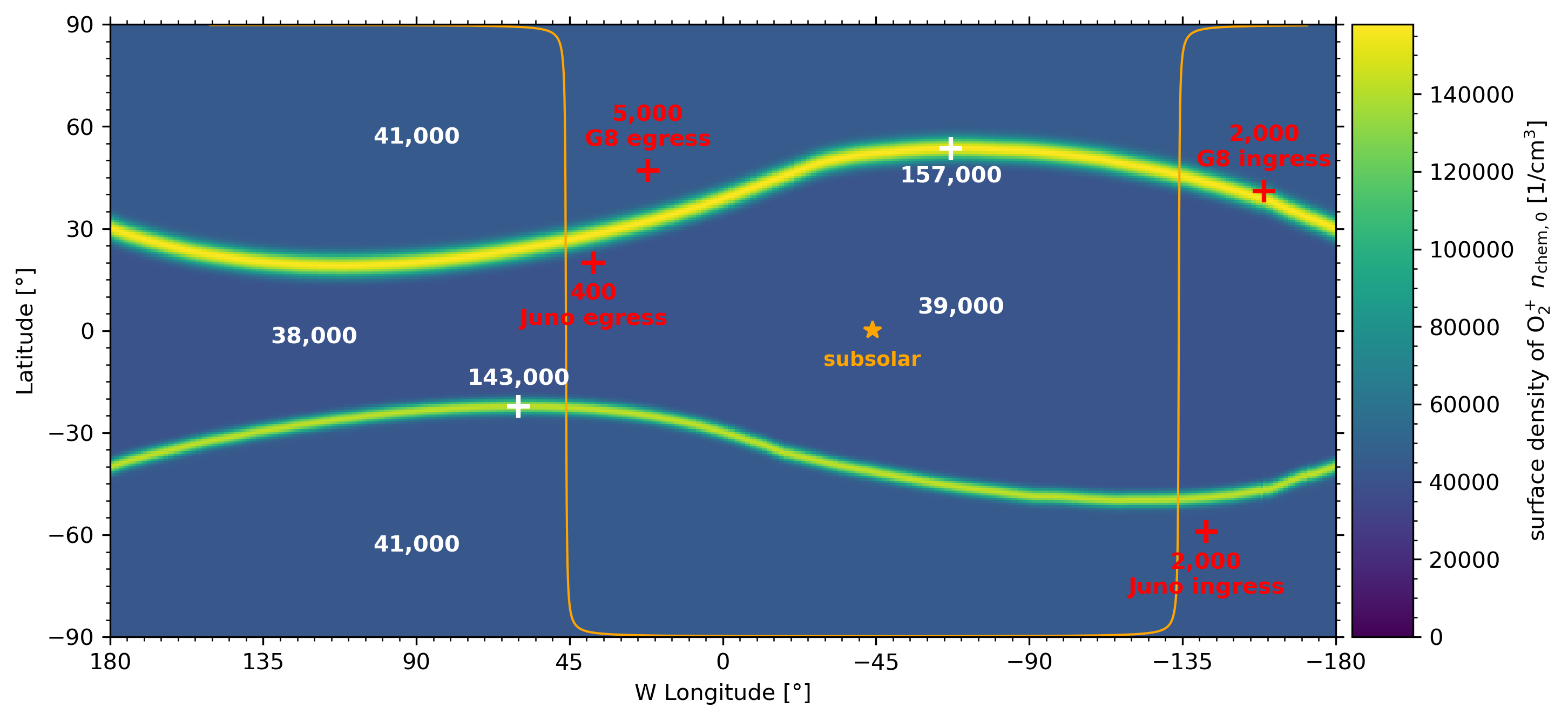}
      \caption{
      O$_2^+$ density at Ganymede's surface, modeled from the chemical equilibrium between ionization and recombination of O$_2$, using the electron-impact ionization rates from Figure \ref{fig_1356_observation_map}c.
      Photoionization is included in the sunlight region, which is shown by the terminator in orange.
      White numbers show modeled density values in cm$^{-3}$.
      For comparison, the red numbers show electron densities detected by Galileo (G8) and Juno radio occultation observations under the assumption of a uniform ionosphere \citep{Kliore1998, Buccino2022}.
      }
      \label{fig_equilibrium_density_map}
\end{figure}

Figure \ref{fig_equilibrium_density_map} shows the resulting O$_2^+$ densities at Ganymede's surface.
A density of $\sim$40,000 cm$^{-3}$ is found in the background of the auroral ovals, while the density peaks at $>10^{5}$ cm$^{-3}$ at the OCFB.
For comparison, the red numbers in Figure \ref{fig_equilibrium_density_map} (values in $\mathrm{cm}^{-3}$) show the locations of electron densities derived from Galileo (G8) and Juno ingress (I) and egress (E) radio occultation observations \citep{Kliore1998,Buccino2022}.
Note that Juno's egress occultation apparently barely missed the northern OCFB and auroral oval location but still detected the lowest density of all the occultation experiments.
Although uncertainties in electron densities from radio occultation experiments are relatively large (30-70\% for total electron content \citep{Buccino2022}), our modeled densities from chemical equilibrium are at least an order of magnitude larger than densities suggested by all previous observations (see Section \ref{sec_introduction}).

There are two possible explanations for this discrepancy.
\begin{enumerate}
\item
All free parameters of Equation \ref{eq_surface_density} are at their unrealistic extremes.
Neglecting photoionization, a surface density of $<5000$ cm$^{-3}$  in the background region from chemical equilibrium between ionization and recombination would require:
      \begin{itemize}
      \item
      An ionization-to-excitation ratio close to its lower limit of $\beta_{\mathrm{ion/1356}}\approx10$, which only occurs for very cold electron distributions ($<20 \;\mathrm{eV}$ monoenergetic or $<5 \;\mathrm{eV}$ Maxwellian).
      For 5 eV Maxwellian electrons, the excitation rate coefficient is low (Figure \ref{fig_cross_section_ratio}c), so that high electron densities ($>100\;\mathrm{cm}^{-3}$) would be necessary.
      \item
      A maximum recombination rate of the produced O$_2^+$ plasma associated with an electron temperature of $\sim$$100 \;\mathrm{K}$.
      \item
      A maximum OI 1356 \r{A} brightness of $\sim$2.5 R in the background region.
      \end{itemize}
\item
Ganymede's ionosphere is not in chemical equilibrium, and transport processes are important.
\end{enumerate}
Because it is implausible that extreme values for multiple parameters would concur, we conclude that Ganymede's ionosphere is dominated by transport.
We will try to further constrain this in the next section.

\subsection{Ionospheric Outflow}
\label{sec_ionospheric_outflow}
Based on the electron-impact ionization rates found in Section \ref{sec_ionization_rates_ganymede}, significant transport processes are necessary to maintain realistic ionospheric densities.
We now evaluate the ratio of O$_2^+$ loss processes in the ionosphere: dissociative recombination and transport.
For simplicity, we only consider vertical plasma flow and calculate the balance between production and loss within a column above Ganymede's surface up to an altitude of $H_{\mathrm{top}}=200$ km, which is ten times the scale height of neutral O$_2$.
Within this column, we adopt an exponential O$_2^+$ plasma density profile with surface density $n_0$ and a constant scale height $H_{\mathrm{e}^-}$,
\begin{equation}
      n\left(h\right) = n_0 \mathrm{e}^{-h/H_{\mathrm{e}^-}}.
\end{equation}

\begin{figure}[htbp]
      \centering
      \includegraphics[width=1.0\textwidth]{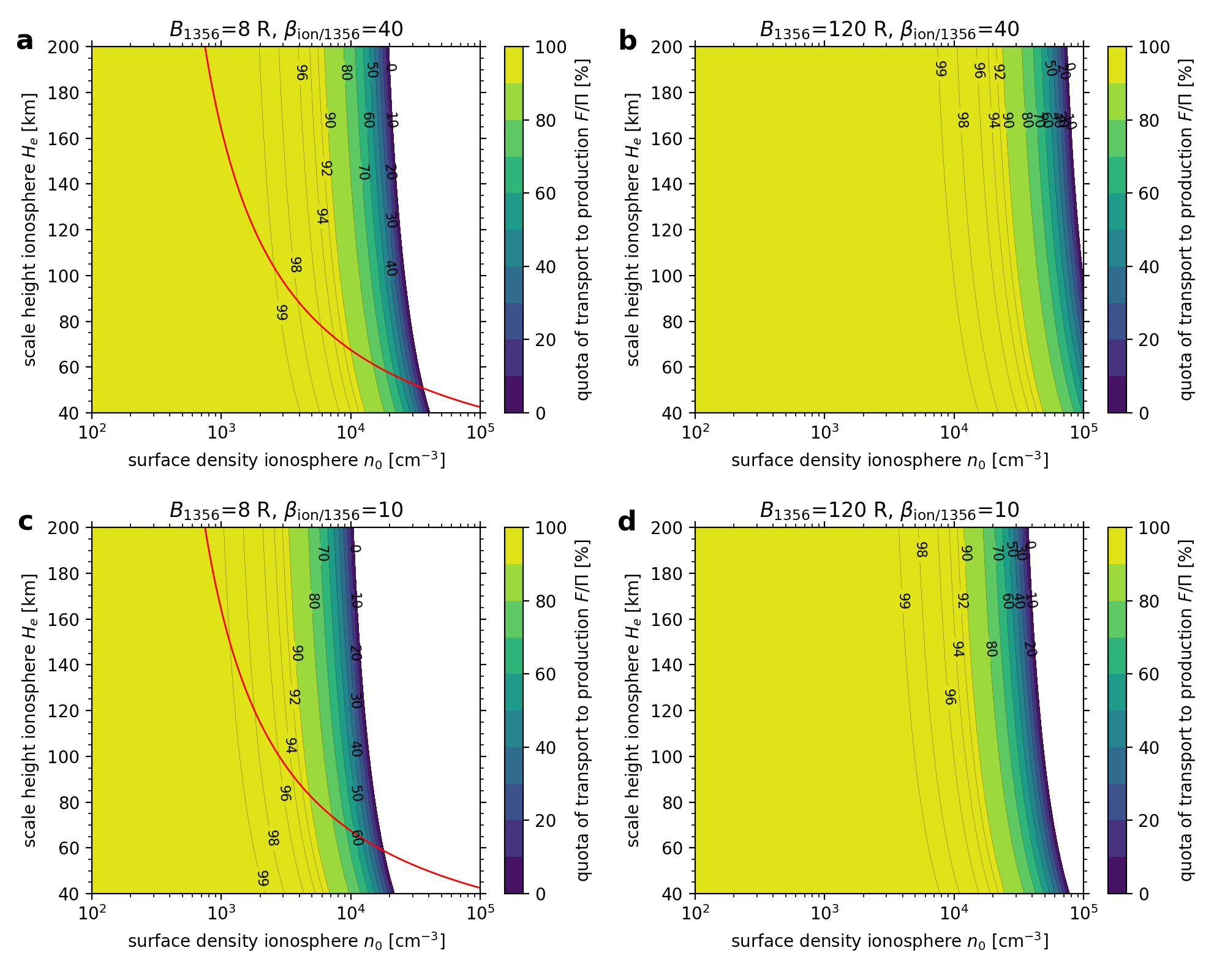}
      \caption{
            Percentages of ion loss by transport.
            The figures show how much of the O$_2^+$ produced by electron-impact and photoionization is transported out of an ionospheric column between Ganymede's surface and 200 km altitude as function of surface densities $n_0$ and scale heights $H_{\mathrm{e}^-}$.
            The top row uses a realistic ionization-to-excitation ratio of $\beta_{\mathrm{ion/1356}}=40$.
            Figure \textbf{a} considers ionization rates consistent with 8 R emissions, which are typical of polar and equatorial regions.
            Figure \textbf{b} considers ionization rates consistent with 120 R emissions, which are typical of the auroral oval.
            The bottom row (Figures \textbf{c}, \textbf{d}) shows calculations for the lower limit of ionization, $\beta_{\mathrm{ion/1356}}=10$.
            For the white parameter space, recombination rates are higher than ionization rates.
            The red line shows density models consistent with in situ measurements during the G2 flyby \citep{Eviatar2001a}.
      }
      \label{fig_outflow_quota}
\end{figure}

For a stationary ionosphere, we integrate Equation \ref{eq_continuity} over the column to find an equation for the particle flux $F = n_\mathrm{top} v_\mathrm{top} - n_0v_0$ onto Ganymede's surface and out of the ionosphere:
\begin{equation}
      F = \Pi\left(1-\mathrm{e}^{-H_\mathrm{top}/H_{\mathrm{e}^-}}\right)
      -\frac{1}{2}\alpha n_0^2H_{\mathrm{e}^-}\left(1-\mathrm{e}^{-2H_\mathrm{top}/H_{\mathrm{e}^-}}\right).
\end{equation}
Here, the velocities $v_\mathrm{top}$ and $v_0$ are defined as positive upwards, and the total column production rate is defined as
\begin{equation}
      \Pi = \Pi_\mathrm{ion}+\Pi_\mathrm{h\nu} =\beta_{\mathrm{ion/1356}}B_{1356} + \nu_{\mathrm{h\nu}}N_{\mathrm{O}_2}.
\end{equation}
The ratio $F/\Pi$ can be used to estimate the loss quota due to transport relative to the total O$_2^+$ production rate.
While the column production rate is well constrained because electron-impact ionization dominates photoionization, the recombination rate depends on the plasma density model.
Thus, $F/\Pi$ varies for different ionospheric scale heights $H_{\mathrm{e}^-}$ and surface densities $n_0$.
Both are rather less constrained.
Figures \ref{fig_outflow_quota}a and \ref{fig_outflow_quota}b show $F/\Pi$ as a function of $H_{\mathrm{e}^-}$ and $n_0$ for average emission brightnesses of 8 R in the background region and 120 R in the auroral ovals.
By including photoionization, we assume a location on the day side, however, as discussed in the previous section, it contributes minimally.
We use the same recombination coefficient for 1000 K electrons and the same O$_2$ column density, as described in Section \ref{sec_density_model}. 
For $\beta_{\mathrm{ion/1356}}$, we again adopt an intermediate value of 40, which aligns with the electron distributions detected by Juno.
As a guideline, the parameter space of density models is indicated in red that is consistent with the only available in situ measurement of 200 cm$^{-3}$ at an altitude of 263 km during the G2 flyby \citep{Eviatar2001a}.
In the white region, i.e., for high surface densities, the associated density model would result in a higher recombination rate than the production rate.

In the background region (Figure \ref{fig_outflow_quota}a), $>$90\% of the produced O$_2^+$ ions must be transported out of the ionosphere to maintain an ionospheric density $<5000$ cm$^{-3}$ at the surface.
If the ionosphere has a small scale height of about 50 km \citep[e.g.][]{Yasuda2024}, the transport quota increases to $>$98\%.
Ensuring consistency with the G2 measurement, recombination is only stronger than transport if the ionosphere has a low scale height of $H_{\mathrm{e}^-}<55$ km, although this requires higher than expected surface densities of $>25,000$ cm$^{-3}$.
Even in case of cold ionizing electrons and the lower limit of $\beta_{\mathrm{ion/1356}}=10$ (Figure \ref{fig_outflow_quota}c), a transport quota of $>$80\% is required to maintain a surface density of $<5000$ cm$^{-3}$.
In all cases, a surface density of $<1000$ cm$^{-3}$ is only possible if $>$99\% of the produced O$_2^+$ is transported out of the ionosphere.

To maintain a surface density of $<10,000$ cm$^{-3}$ for ionization rates consistent with the $\sim$120 R emissions in the auroral ovals, the transport quota must be even higher (Figures \ref{fig_outflow_quota}b, d).
In this case, a transport quota of 90\% results in a surface density ranging from  10,000 cm$^{-3}$ to 50,000 cm$^{-3}$.

Considering the global electron-impact ionization rate of $1.3-7.6\times10^{26}\;\mathrm{s}^{-1}$ (Section \ref{sec_ionization_model}) and the high likelihood of $>$90\% loss due to transport, we estimate a lower limit of $1.2-6.8\times10^{26}\;\mathrm{s}^{-1}$ for the O$_2$ surface release rate to maintain Ganymede's atmosphere.
This rate is consistent with the total release rate from the modeling of \cite{Vorburger2024}, although, besides ionization, they also considered additional O$_2$ losses from gravitational escape, surface adsorption, and dissociation.

Thus far, we have only estimated the total O$_2^+$ loss rate due to transport processes.
To further assess the ratio of upward and downward fluxes - i.e., collision with Ganymede's surface and ionospheric outflow into the magnetosphere - we additionally incorporated upward electron and ion fluxes detected by JADE during Juno's flyby.
Using the electron intensities measured when Juno was on open magnetic field lines, we calculated the energy spectrum of electrons within the upward loss cone of $>$165° pitch angle \citep{Allegrini2022a}.
Integrating the spectrum yields an upward electron flux of $8\times10^6$ cm$^{-2}$s$^{-1}$.
However, the maximum of the spectrum is below the detection threshold of 100 eV, meaning that most of the outflowing electrons are cold and likely undetected.
Thus, the observed electron flux can only be interpreted as a lower limit of the ionospheric outflow.
Unlike the electrons, the energy distribution of O$_2^+$ ions, as shown in \cite{Valek2022}, is mostly covered by the detection limits.
According to their analysis, the upward fluxes of O$_2^+$ on open magnetic field lines were $>0.1\times10^{8}$ cm$^{-2}$s$^{-1}$ and peaked at $0.5\times10^{8}$ cm$^{-2}$s$^{-1}$ shortly after reaching the closest approach at an altitude of 1046 km.
Scaled to an altitude of 200 km, this corresponds to an observed ionospheric outflow flux of $0.2-0.8\times10^{8}$ cm$^{-2}$s$^{-1}$.
Considering the O$_2^+$ production rate of $3.2\times10^{8}$ cm$^{-2}$s$^{-1}$ for 8 R of OI 1356 \r{A} emissions and a transport quota of 90\%, this means that approximately 10-30\% of the loss due to transport is through ionospheric outflow, while 70-90\% are lost at Ganymede's surface.
This ratio is consistent with the O$_2^+$ escape rate of 19\% found by previous Monte Carlo modeling of the ionosphere \citep{Carnielli2019}, although the global production rate they employed is more than one order of magnitude lower than that derived from our analysis.

Due to their heavy mass, ionospheric outflow of O$_2^+$ ions contributes the most to the mass loading of Ganymede's magnetosphere.
Applying the outflow quota to the global ionization rate yields an outflow rate of $0.1-2\times10^{26}\;\mathrm{s}^{-1}$ and a mass loading rate of $0.5-11\;\mathrm{kg}\;\mathrm{s}^{-1}$.

\section{Discussion}
\label{sec_discussion}

\subsection{Precipitating Electron Density and Energy Fluxes}

\begin{figure}[htbp]
      \centering
      \includegraphics[width=0.5\textwidth]{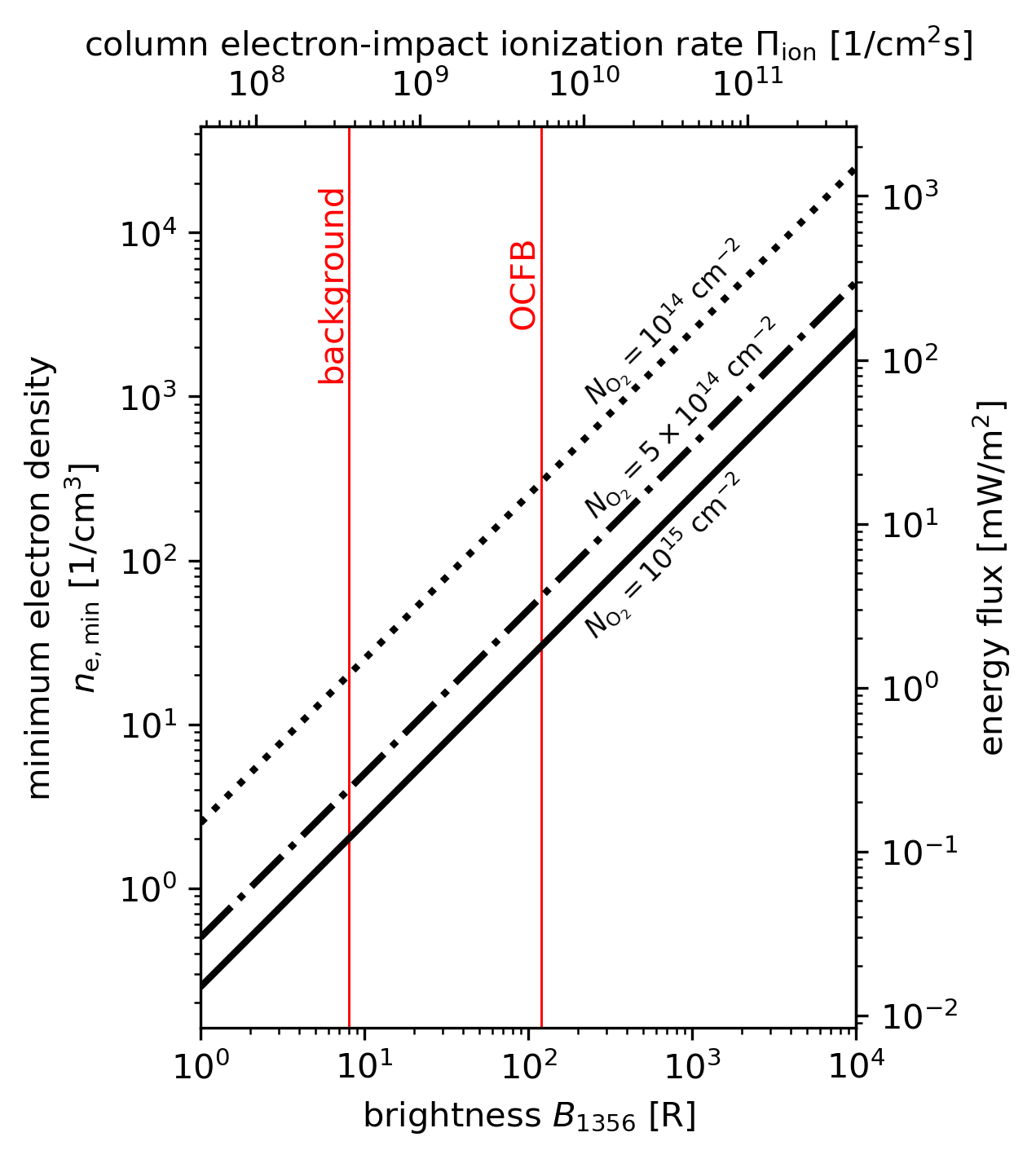}
      \caption{
      Lower limits of electron density and downward energy fluxes required to excite O$_2$ OI 1356 \r{A} emissions of certain brightnesses.
      The maximum possible excitation rate coefficient of a monoenergetic distribution at 185 eV was used for calculation ($k_{\mathrm{1356,max}}=4\times10^{-9}$ cm$^3$s$^{-1}$).
      Associated column electron-impact ionization rates were calculated with $\beta_{\mathrm{ion/1356}}=46$, which corresponds to 185 eV electrons.
      The three black lines consider different O$_2$ column densities.
      The red lines indicate the average brightnesses in the auroral oval (OCFB) and in the background region.
      }
      \label{fig_min_electron_density}
\end{figure}

We estimate the flux of precipitating electrons required to produce the ionization and excitation rates found by our analysis.
Because the excitation rate coefficient (Figure \ref{fig_cross_section_ratio}c) has a maximum, $k_{\mathrm{1356,max}}$, a lower limit of electron density, $n_{\mathrm{e,min}}$, can be evaluated to produce observed OI 1356 \r{A} emission brightnesses:
\begin{equation}
      n_{\mathrm{e,min}} > \frac{B_\mathrm{1356}}{N_{\mathrm{O}_2} k_{\mathrm{1356,max}}}.
\end{equation}
For Maxwell distributions, the maximum rate coefficient is $3.5\times10^{-9} \;\mathrm{cm}^3 \mathrm{s}^{-1}$ and corresponds to an electron temperature of 220 eV.
Monoenergetic electrons excite most efficiently at $\sim$185 eV, where the maximum rate coefficient is $4\times10^{-9} \;\mathrm{cm}^3 \mathrm{s}^{-1}$.
Assuming electrons with this energy, Figure \ref{fig_min_electron_density} shows the lower limit of electron density for three selected O$_2$ column densities to produce a certain emission brightness.
The upper axis associates an electron-impact ionization rate with the brightness values using $\beta_{\mathrm{ion/1356}}=46$, which corresponds to 185 eV electrons, although ionization maximizes at a slightly higher electron energy of $\sim$250 eV.
Required energy fluxes in the downward direction are estimated as 
\begin{equation}
      \frac{1}{8}n_em_e\int_{0}^{10^5\mathrm{ eV}} v^3f(v) \,\mathrm{d}v.
\end{equation}
For monoenergetic distributions, energy fluxes are calculated as $n_eW_e\sqrt{W_e/8m_e}$ and are shown on the right axis.
Note that these densities and fluxes are the lower limits for monoenergetic electrons.
However, since the maximum rate coefficient for Maxwellian electrons is reduced by only 12.5\%, the lower limits for Maxwellian electrons are only slightly smaller. 

An average brightness of 8 R in the polar and equatorial regions requires at least 2-20 cm$^{-3}$ electrons, depending on the neutral column density.
Using $N_{\mathrm{O}_2} = 5\times10^{14} \;\mathrm{cm}^{-2}$, the lower limit of 4 cm$^{-3}$ electrons is associated with an energy flux of $\sim0.25\;\mathrm{mW}/\mathrm{m}^2$.

To excite the bright emissions of the auroral ovals (averaging 120 R), a lower limit of 30 cm$^{-3}$ electrons is required.
This is approximately six times higher than typical upstream plasma densities in the Jovian magnetosphere \citep{Kivelson2004}.
In the case of a thin atmosphere with $N_{\mathrm{O}_2} = 10^{14} \;\mathrm{cm}^{-2}$, even 300 cm$^{-3}$ electrons are required.
For an atmosphere with $N_{\mathrm{O}_2} = 5\times10^{14} \;\mathrm{cm}^{-2}$, a minimum electron density of 60 cm$^{-3}$ and $\sim3.6\;\mathrm{mW}/\mathrm{m}^2$ is required to generate the emissions observed in the auroral ovals, still assuming monoenergetic electrons.

For the electron distributions detected by Juno near Ganymede \citep{Ebert2022a}, we calculate a rate coefficient of $k_{\mathrm{1356}}=1-3\times10^{-9}\;\mathrm{cm}^3 \mathrm{s}^{-1}$ with additional uncertainty due to the incomplete energy spectra (see Section \ref{sec_ionization_rates_from_emissions}).
Assuming these distributions and an intermediate dense atmosphere, $80-240 \;\mathrm{cm}^{-3}$ electrons are required to excite 120 R emissions.
Compared to monoenergetic or Maxwellian electrons, the energy flux associated with the detected distributions is higher due to their energetic tail.
The minimum downward energy flux required to excite 120 R with these electrons is $20-100\;\mathrm{mW}/\mathrm{m}^2$.

\subsection{Temporal and Spatial Variability}
\label{sec_variability}
Our model accounts for large-scale spatial structures of auroral emissions and ionization rates, considering Gaussian bands close to the OCFB and constant equatorial and polar background regions.
Due to the change of the Jovian magnetic field with Ganymede's position relative to the magnetospheric current sheet, the OCFB location varies with Jupiter's synodic rotation ($\sim$10.54h) \citep{Saur2015,Musacchio2017}.
The magnetosphere model that was used to predict the location of the OCFB at the time of Juno's flyby \citep{Duling2022} can be used to adapt the OCFB to other times.

Acceleration due to magnetic reconnection is a promising candidate for the origin of exciting electrons \citep{Ebert2022a,Joseph2024,Kaweeyanun2020}.
However, the exact generation mechanism of the auroral excitation remains a subject of ongoing research.
It is likely that the mechanism is spatially and temporally sub-structured, which leads to variability of the precipitating electron fluxes.
This presumably causes variations in auroral brightness and the width of the auroral bands.
Both HST and Juno observations suggest spatial brightness variations with longitude \citep{Marzok2022a,Greathouse2022}.
Short-time brightness variability has yet to be identified in auroral observations \citep{Saur2022}.
However, there are indications of brightness variations with the synodic period.
HST observations revealed a north-south asymmetry from Ganymede's position below or above the current sheet \citep{Saur2022,Milby2024}.

The variations with the synodic period may be superimposed by dynamics of Ganymede's magnetosphere on a shorter timescale.
Suggested Kelvin-Helmholtz instabilities \citep{Stahl2023} or flux ropes \citep{Zhou2020} may affect the magnetic field geometry, which guides the precipitating electrons.
These dynamical processes may lead to short-time variations in the location of the auroral ovals, resulting in blurring in longer-term observations.
Unfortunately, Juno's one-time 12-minute UVS observation cannot distinguish between spatial and temporal dependencies.
Therefore, it remains unclear whether the auroral bands are several degrees wide, as our analysis suggests and our model assumes, or if they are sharply localized but blurred in the observation.
This can eventually be resolved by future missions or a better understanding of the precipitating electron fluxes.

As the characteristics of these variations are complex and not yet fully understood, they are not included in our analysis.
Instead, this work's modeling results are quantitative estimates of globally averaged electron-impact ionization rates on Ganymede.

\subsection{Scale Height of the O$_2^+$ Ionosphere}
\label{sec_discussion_tec_altitude}
In Section \ref{sec_density_model}, we compare modeled ionospheric densities with those expected from radio occultation experiments to conclude that transport is of significant importance for Ganymede's ionosphere.
To evaluate the uncertainty of this comparison, we briefly discuss the capability of the radio occultation experiments to detect the ionospheric density close to Ganymede's surface.
The loss of the radio link between ingress and egress occultations prevents total electron content measurements of beams directly tangential to Ganymede's surface, i.e., those passing through the lowest layer of the ionosphere. 
For Juno, a "snap-lock" technique was used to relock the radio signal in less than one second, enabling measurements down to $\sim$10 km altitude \citep{Buccino2022}.
For the chemical equilibrium model (Equation \ref{eq_plasma_density}), the ionospheric density scales with the double neutral scale height.
Using $H=20$ km, the density at 10 km altitude is thus $0.78$ times lower than at the surface.

However, the ionosphere was found to be not in chemical equilibrium, so that transport processes significantly increase its scale height.
This reduces the difference between the ionospheric densities at 10 km altitude and at the surface.
Consequently, the density at the surface is closer to the radio occultation measurements than it would be in the case of chemical equilibrium.
At higher altitudes, where the neutral density and ionization rates are lower, transport has an even stronger effect on the ionospheric density.
This suggests a non-constant O$_2^+$ scale height that increases with altitude, though the exact dependency is unknown.

\subsection{Evolution of Surface Ice}
As atmospheric O$_2$ is primarily produced by radiolytic dissociation of H$_2$O, our derived production and loss rates for O$_2^+$ ions can be used as tracer for the evolution of Ganymede's surface ice \citep{Szalay2024}.
Following our conclusion that 90\% of the $1.3-7.6\times10^{26}\;\mathrm{s}^{-1}$ O$_2^+$ production is lost through transport and at most 10\% recombine to neutral oxygen, $1.2-6.8\times10^{26}\;\mathrm{s}^{-1}$ of O$_2$ must be released from the surface to sustain the atmosphere.
Assuming that all released O$_2$ originates from radiolytic dissociation of H$_2$O, which yields H$_2$ and O$_2$ in a 1:0.5 ratio, the rate of radiolysis would be $2.3-13.7\times10^{26}\mathrm{s}^{-1}$.

However, we estimate that $70-90$\% of the transport loss of O$_2^+$, $0.8-6.2\times10^{26}\;\mathrm{s}^{-1}$ or $4-33$ kg s$^{-1}$, is due to impact on the surface, where it rapidly recombines to neutral O$_2$.
Since the lighter H$_2$ escapes more readily, the surface impact quota is likely much smaller than that of O$_2$, though a detailed quantification is beyond the scope of this work.
This imbalance may lead to preferential loss of hydrogen, resulting in net oxygenation of the surface.
The accumulated oxygen can subsequently be re-released into the atmosphere, establishing a pure O$_2$ - O$_2^+$ cycle that would substantially reduce the required rate of ice dissociation.

Nevertheless, a lower limit for the dissociation rate of surface ice can be inferred from the ionospheric outflow of O$_2^+$, which is 10-30\% of the total transport loss.
The O$_2^+$ outflow rate of $0.1-2\times10^{26}\;\mathrm{s}^{-1}$ corresponds to an H$_2$O loss rate of $0.2-4\times10^{26}\;\mathrm{s}^{-1}$, or $0.7-12$ kg s$^{-1}$, equivalent to an erosion of $0.03-0.5$ cm Myr$^{-1}$ for Ganymede's surface.
These erosion rates are somewhat lower than those estimated for Europa, where $1.5\pm0.8$ cm Myr$^{-1}$ has been calculated \citep{Szalay2024}.
However, since Ganymede's surface area is three times larger than Europa's, the total release rate from radiolysis of water ice is comparable to that at Europa ($13\pm7$ kg s$^{-1}$).


\section{Summary}
\label{sec_summary}
Owing to the similar energy dependence of the relevant cross sections, we found that electron-impact ionization rates of O$_2$ are well constrained by observable OI 1356 \r{A} emissions.
This approach eliminates the need for prior knowledge of precipitating electron fluxes and neutral densities and avoids the uncertainties associated with estimating them.
We applied this new method to Juno UVS observations of Ganymede's aurora and derived the following conclusions on the emission and ionization characteristics:
\begin{itemize}
      \item The ratio of electron-impact ionization to OI 1356 \r{A} excitation rates of O$_2$ consistently falls within the range of 10 to 60.
      \item Further constraints can be inferred for electrons with Maxwellian energy distribution. An ionization-to-excitation ratio of 10 implies $T<4$ eV and requires 13 times more electrons than a distribution with $T\approx12$ eV, which yields a ratio of 20. For Maxwellian electrons the ratio is thus realistically confined to 20-60.
      \item For Kappa distributions with a thermodynamic kappa index close to 3/2 the ionization-to-excitation ratio is confined to 30 to 50.
      \item Measured distributions of electrons precipitating to Ganymede indicate an ionization-to-excitation ratio of 32 to 47, which is our recommended range.
      \item Ganymede's auroral ovals in the OI 1356 \r{A} doublet are well described by 3-5° latitude wide Gaussian distributions centered on the OCFB with an average peak brightness of $\sim$120 R.
      \item The mean OI 1356 \r{A} brightness in the equatorial and polar regions is $\sim$8 R.
      \item Electron-impact ionization rates are at least an order of magnitude higher than photoionization rates at Ganymede.
      \item We provide an empirical global map with ionization rates of O$_2$ at Ganymede that can be used in further modeling.
      \item Average column ionization rates are $\sim5\times10^{9}$ cm$^{-2}$s$^{-1}$ at the OCFB and $\sim3\times10^{8}$ cm$^{-2}$s$^{-1}$ in the equatorial and polar regions.
      \item The total global ionization rate at Ganymede is $1.3-7.6\times10^{26}$ s$^{-1}$.
      \item Transport processes are the dominant loss mechanism in Ganymede's O$_2^+$ ionosphere with 60-80\% by surface impact, 10-30\% by ionospheric outflow and at maximum 10\% by dissociative recombination.
      \item The ionospheric outflow of O$_2^+$ is $0.1-2\times10^{26}\;\mathrm{s}^{-1}$ or $0.5-11\;\mathrm{kg}\;\mathrm{s}^{-1}$.
      \item The erosion of Ganymede's surface from radiolytic dissociation of water ice is 0.03-0.5 cm Myr$^{-1}$
\end{itemize}

\section*{Conflict of Interest}
The authors declare there are no conflicts of interest for this manuscript.

\section*{Open Research Section}
The cross sections and rate coefficients of electron-impact ionization and OI 1356 \r{A} excitation, and their ratios, as used in this study and shown in Figure \ref{fig_cross_section_ratio} are available at a Zenodo repository via \url{https://doi.org/10.5281/zenodo.17976287} with CCA 4.0 license \citep{Duling2025b}.
Juno UVS data can be obtained from the Planetary Data System (PDS) at \url{https://pds-atmospheres.nmsu.edu/data_and_services/atmospheres_data/JUNO/uvs.html}.
The calibrated data \citep{Trantham2014} used in this study is located at \url{https://pds-atmospheres.nmsu.edu/PDS/data/jnouvs_3001/}.
The maps of OI 1356 \r{A} and OI 1304 \r{A} emission brightness, shown in Figure \ref{fig_1356_observation_map}, as well as the corresponding estimations of reflected sunlight and the Juno UVS exposure times are available at a Zenodo repository via \url{https://doi.org/10.5281/zenodo.17964985} with CCA 4.0 license \citep{Duling2025a}.
The modeled location data of the OCFB during Juno's flyby used in this study are available at a Zenodo repository via \url{https://doi.org/10.5281/zenodo.7096938} with CCA 4.0 license \citep{Duling2022a}.
The fitted location data of the OCFB during Juno's flyby, as shown in Figure \ref{fig_1356_observation_map}b, are available at a Zenodo repository via \url{https://doi.org/10.5281/zenodo.17963354} with CCA 4.0 license \citep{Duling2025}.

\acknowledgments
S.D. and J.S. received funding from the European Research Council (ERC) under the European Unions Horizon 2020 research and innovation programme (grant agreement No. 884711).
T.G. was funded by the NASA's New Frontiers Program for Juno via contract NNM06AA75C with the Southwest Research Institute.

%
%

\bibliography{ExoOceans}

%
%
%
%
%

\end{document}